\documentclass{aa}
\usepackage[varg]{txfonts}
\usepackage{graphicx}
\graphicspath{{figures/}}
\usepackage{natbib}
\bibpunct{(}{)}{;}{a}{}{,} 
\usepackage{hyperref}
\hypersetup{
	colorlinks=true,
	linkcolor=blue,
	citecolor=blue
}

\begin{document}

\title{GAMA+KiDS: Alignment of galaxies in galaxy groups and its dependence on galaxy scale}

\author{Christos Georgiou \inst{1}\thanks{georgiou@strw.leidenuniv.nl}
	\and Nora Elisa Chisari \inst{2}
	\and Maria Cristina Fortuna \inst{1}
	\and Henk Hoekstra \inst{1}
	\and Konrad Kuijken \inst{1}
	\and Benjamin Joachimi \inst{3}
	\and Mohammadjavad Vakili \inst{1}
	\and Maciej Bilicki \inst{4}
	\and Andrej Dvornik \inst{1}
	\and Thomas Erben \inst{5}
	\and Benjamin Giblin \inst{6}
	\and Catherine Heymans \inst{6}
	\and Nicola R. Napolitano \inst{7}
	\and HuanYuan Shan \inst{8}}
\institute{Leiden Observatory, Leiden University, Niels Bohrweg 2, 2333 CA Leiden, The Netherlands
	\and Department of Physics, University of Oxford, Keble Road, Oxford OX1 3RH, UK
	\and Department of Physics and Astronomy, University College London, Gower Street, WC1E 6BT London, UK
	\and Center for Theoretical Physics, Polish Academy of Sciences, al. Lotnik\'ow 32/46, 02-668, Warsaw, Poland
	\and Argelander-Institut für Astronomie, Auf dem Hügel 71, 53121 Bonn / Germany
	\and Institute for Astronomy, University of Edinburgh, Royal Observatory, Blackford Hill, Edinburgh, EH9 3HJ, UK
	\and School of Physics and Astronomy, Sun Yat-sen University Zhuhai Campus, 2 Daxue Road, Tangjia, Zhuhai, Guangdong, 519082, P.R. China
	\and Shanghai Astronomical Observatory (SHAO), Nandan Road 80, Shanghai 200030, China}

\date{Received <date> / Accepted <date>}

\abstract{Intrinsic galaxy alignments are a source of bias for weak lensing measurements as well as a tool for understanding galaxy formation and evolution. In this work, we measure the alignment of shapes of satellite galaxies, in galaxy groups, with respect to the brightest group galaxy (BGG), as well as alignments of the BGG shape with the satellite positions, using the highly complete Galaxy And Mass Assembly (GAMA) spectroscopic survey and deep imaging from the Kilo Degree Survey. We control systematic errors with dedicated image simulations and measure accurate shapes using the DEIMOS shape measurement method. We find a significant satellite radial alignment signal, which vanishes at large separations from the BGG. We do not identify any strong trends of the signal with galaxy absolute magnitude or group mass. The alignment signal is dominated by red satellites. We also find that the outer regions of galaxies are aligned more strongly than their inner regions, by varying the radial weight employed during the shape measurement process. This behaviour is evident for both red and blue satellites. BGGs are also found to be aligned with satellite positions, with this alignment being stronger when considering the innermost satellites, using red BGGs and the shape of the outer region of the BGG. Lastly, we measure the global intrinsic alignment signal in the GAMA sample for two different radial weight functions and find no significant difference.}

\keywords{galaxies: evolution - large-scale structure of Universe - gravitational lensing: weak - cosmology: observations}

\titlerunning{Alignment of galaxies in galaxy groups and its dependence on galaxy scale}
\maketitle

\section{Introduction}
\label{sec:Introduction}

One of the major unsolved problems of modern cosmology, and the established concordance model, dubbed $\Lambda$CDM, entails the mystery around dark matter and dark energy. These components account for approximately 95\% of the energy density of the universe today, with their presence established through several observations \citep[e.g.][]{SNIa, BOSSclustering, Planck} but their nature remaining elusive. This is mostly due to the exceptionally low, if not null, cross-section for interactions involving the non-baryonic particles of dark matter \citep[see e.g.][for a review]{DMReview} as well as the difficulty in observationally obtaining enlightening information about dark energy \citep[for a review, see][]{DEreview}. 

In order to shed light onto the nature of these "dark" components of the universe we need to use robust and complementary observations that are sensitive to their effects. One observational technique of particular interest is weak gravitational lensing, the coherent distortion of light bundles that travel through a matter-bent spacetime \citep[see e.g.][]{LensingReview1, LensingReview2}. Lensing is sensitive to the intervening matter (baryonic and non-baryonic) between the light source and the observer as well as the geometry of the universe, thus serves as a great tool for constraining both dark matter and dark energy. 

Weak gravitational lensing measurements use shape correlations of galaxies in order to extract the lensing signal and infer cosmological parameters \citep[for a review see][]{CosmicShearReview, GGLReview}. However, lensing measurements are sensitive to certain physical and observational systematic uncertainties that need to be taken into account. One major astrophysical contaminant is the alignment of galaxies with other galaxies and the matter density field, known as intrinsic galaxy alignment \citep[see e.g.][for a review]{JoachimiReview}. Intrinsic alignments produce shape correlations between physically associated galaxies (named intrinsic-intrinsic or II), as well as the more complicated correlation between background and foreground galaxy shapes (named gravitational-intrinsic or GI correlation). These correlations, if ignored, can significantly bias lensing measurements \citep[e.g.][]{KirkReview, TroxelReview}.

Intrinsic alignments are commonly mitigated by marginalising over modelled parameters describing them, when extracting lensing information. The most common approach is to use the linear alignment model \citep{LA}, which assumes that galaxy alignments are linearly related to the gravitational field, while often replacing the linear matter power spectrum with its non-linear counterpart, dubbed non-linear linear alignment model \citep[NLA,][]{NLA}. This model has been successful in describing direct measurements of intrinsic alignments on scales down to a few Mpc$/h$ \citep[e.g.][]{Singh1}, and has been used in many studies measuring lensing from large-scale structure \citep[cosmic shear, e.g.][]{KV450, DEScosmicshear, HSCcosmicshear} or combining it with weak lensing from individual lens galaxies \citep[galaxy-galaxy lensing, e.g.][]{Edo3x2, Joudaki, DES3x2}.

For smaller scales, the linear alignment model fails to reproduce observations since the physical mechanism that aligns galaxies at such scales is fundamentally different \citep{Pereira}. Consequently, devising a descriptive model for intrinsic alignments at these scales is challenging, yet crucial to using these information-rich scales in lensing measurements. One approach is the halo model, which describes how galaxies (central and satellites) populate a dark matter halo given its mass, as well as how they are oriented \citep{HaloModel}. This model is the only description of the alignment signal at small scales so far, and can in principle be extended to describe extra dependencies of the signal. While a general consensus for the intrinsic alignment signal on small scales has not yet been reached, it is important to study the phenomenon directly on intra-halo scales and provide insight for building an accurate halo model.

Intrinsic alignments have been studied in many cosmological and numerical simulations, both on large and intra-halo scales \citep[e.g.][]{projSubhalos, dmandgas, Tenneti, Marco, Elisasim, Hilbert}. A strong alignment signal on small scales is seen in these simulations, with satellite galaxies being radially aligned with their host halo centres. However, the picture is not very clear in observations, with many studies finding satellite orientations consistent with random \citep[e.g.][]{Schneider, Elisagroups, Sifon2015} while others see a positive radial alignment signal \citep[e.g.][]{IAlens,Huang2}. 

This discrepancy can be attributed to a difference in galaxy populations, but also to a scale dependence of the alignment signal, i.e. outer galaxy regions exhibiting a stronger alignment than inner ones. Such a dependence is expected due to the tidal nature of the phenomenon and has been seen in hydrodynamical simulations, where the alignment signal was stronger when the method used to define galaxy shapes up-weighted outer regions of galaxies \citep{Marco, Elisasim, Hilbert}. Observations have also hinted towards this, with \citet{Singh2, Huang2} looking into alignments as measured by different shape measurement methods and finding stronger alignments for methods more sensitive to the outer shape of a galaxy. In addition, \citet{Georgiou} found that the alignment signal changes on small scales when shapes were measured from different broad band filters, with the change mostly sourced by satellite galaxies. Since galaxies exhibit colour gradients, certain regions of galaxies can be more prominent in one broad band filter observation than another, hinting towards a scale dependence of the alignment signal.

In this work we measure the alignment of shapes of satellite galaxies with respect to their group's centre (traced by the brightest group galaxy, or BGG), as well as the alignment of shapes of BGGs with the positions of satellites. We use galaxy groups from the Galaxy And Mass Assembly survey\footnote{http://www.gama-survey.org} \citep[GAMA;][]{Driver2009,Driver2011,Liske2015, GAMADR3} and apply the DEIMOS shape measurement method \citep{DEIMOS} on imaging data from the Kilo Degree Survey\footnote{http://kids.strw.leidenuniv.nl/} \citep[KiDS;][]{deJong2015,deJong2017,KiDSDR4}. Possible trends with observed wavelength, magnitude, group mass and galaxy colour are examined. We also look into the galaxy scale dependence of these alignments by varying the radial weight function employed when measuring galaxy shapes and comparing the resulting alignment signal. This will shed light onto the discrepancy of alignments between observations and simulations, as well as allow for better calibration of the halo model for intrinsic alignments in the future. Lastly, we look at the global intrinsic alignment signal and its dependence on galaxy scale.

This paper is organised as follows: in Sect. \ref{sec:Data} we describe the data used, and the methodology for obtaining a robust alignment signal is outlined in Sect. \ref{sec:Methodology}. Satellite galaxy alignments are presented in Sect. \ref{sec:Results}, BGG alignments with its satellites in Sect. \ref{sec:BGGsat}, and the scale dependence of central galaxies is shown in Sect. \ref{sec:centrals}. Conclusions follow in Sect. \ref{sec:Conclusions}. Throughout this paper we use flat $\Lambda$CDM cosmology with $h=0.7$ and $\Omega_{\rm m}=0.315$.

\section{Data}
\label{sec:Data}

Measuring the alignment of galaxies in galaxy groups requires a robust group catalogue and good imaging data. The former is obtainable from a galaxy sample with accurate redshifts (usually obtained spectroscopically) as well as a highly complete and deep sample, with which we can identify as many group members as possible. The latter can be acquired from several deep imaging surveys that cover large parts of the sky. In this work, we use data from the GAMA survey, combined with the deep and high quality imaging data of the KiDS survey for highly accurate shape measurements. 

\subsection{Galaxy group sample}
\label{sec:groups}

Our galaxy group sample is drawn from the final release of the GAMA survey \citep{Liske2015, GAMADR3}. The survey acquired spectroscopy for $\sim$300,000 galaxies across a total $\sim$286 deg$^2$ of sky. The survey is split into 5 patches of roughly equal area, three of them in the equatorial regions (named G09, G12 and G15) and two in the southern celestial hemisphere (G02 and G23). Region G02 does not overlap with our imaging data while region G23 has a slightly different target selection and, to avoid complications that may arise due to this difference, we restrict our analysis to the three equatorial regions. That leaves us with $\sim$180 deg$^2$ of area and $\sim$180,000 galaxies. The limiting magnitude for the equatorial regions is $r<19.8$ mag.

The unique characteristic of the survey is its high completeness; in the equatorial regions, redshift information is obtained for 98.5\% of the target galaxies. This allows for a robust estimation of galaxy alignments on small scales, where complications such as fiber collisions are not present. In addition, the completeness enables the construction of a high fidelity group catalogue \citep{Aaron}.

The group catalogue was created using a friends-of-friends grouping algorithm of variable linking length, and contains $\sim$23,600 galaxy groups (with at least two member galaxies) in the equatorial regions, for a total of $\sim$75,000 group galaxies. The redshifts of these groups extend to $z\simeq0.6$ with a median group redshift of $z_{\rm med}=0.213$, which is practically the same as the parent sample. The grouping algorithm has been tested and calibrated using mock catalogues from semi-analytic galaxy models applied to N-body simulations, specifically designed to capture the properties of galaxies in the GAMA survey \citep[see][for details]{Aaron}. These mocks also enable the calculation of unbiased group luminosities, which we make use of in Sec. \ref{sec:Results}. 

We calculate the alignment of satellite galaxies with respect to the brightest group galaxies, which may not always coincide with the group's dark matter halo. This mis-centring was examined in \citet{Violaetal} for GAMA groups where different group centre definitions were considered. It was found that consistent results were produced using the BGG and using the centre as defined iteratively by removing the galaxy further away from the group's centre of light. In this work, for simplicity, we use the BGG as a proxy for the group's centre.

\subsection{Galaxy shape measurements}
\label{sec:shapes}

A robust shape measurement requires high quality, deep imaging data as well as an accurate shape measurement method. We use imaging data from KiDS \citep{deJong2015, deJong2017, KiDSDR4}, a survey specifically designed and optimised for weak gravitational lensing science. The survey aims to cover 1,350 deg$^2$ in the four optical SDSS-like broad bands $u,g,r$ and $i$, down to limiting magnitude of 24.3, 25.1, 24.9 and 23.8 (5$\sigma$ in a 2 arcsec aperture), respectively. This results in deep, high quality images, with a small, nearly uniform Point-Spread function (PSF) \citep{Kuijken}.

The shape measurement method we choose is DEIMOS \citep{DEIMOS}. This is a moment-based method, extracting shape information using moments as measured directly from the image. It is an improvement over other similar techniques  \cite[e.g.][]{KSB, regauss} because it allows usage of high-order moments in correcting for the employed radial weight function, necessary when dealing with noisy imaging data. In addition, the PSF is treated in a fully analytical way, without any prior assumptions on its properties; the limiting factor is how well one can model the PSF on the positions of galaxies. Finally, unlike model-dependent shape measurement techniques \citep[e.g.][]{lensfit}, we can directly vary the radial weight function, which translates to measuring galaxy shapes sensitive to different galaxy regions. A larger weight function will be more sensitive to outer galaxy regions, and using this we can probe the galaxy scale dependence of the alignment signal. 

The PSF used in this work is modelled using orthogonal shapelets \citep{shapelets}, which are Hermite polynomials multiplied by Gaussian functions. Shapelets can be linearly combined (to arbitrary extent) to describe image shapes. This model has been shown to adequately describe the PSF variations in KiDS images \citep{Kuijken}. The weight function is a 2D Gaussian, initially circular, with the scale of the Gaussian being a free parameter called $r_{\rm wf}$. The weight function is then matched iteratively to the ellipticity and centroid of each galaxy while preserving its area to the initial circular function. In order to mask neighbouring galaxies, segmentation maps from \textsc{SExtractor} are used that identify pixels associated with other sources. These pixels are then replaced with random Gaussian noise of the same level as the image. The process followed for the shape measurement is described in detail in \citet{Georgiou}, where characteristics such as the galaxy ellipticity distribution and signal-to-noise can also be seen.

In this work, we measure shapes of galaxies from $r$-band imaging data of KiDS Data Release 4 (DR4), using five different radial weight function sizes (see Sec. \ref{sec:weightfunction}). In order to probe differences of the alignment with the observed wavelength, we use shapes measured in $g,r$ and $i$-band with a single weight function from \citet{Georgiou}, which were based on KiDS Data Release 3 (DR3) and covers fully the GAMA equatorial fields as well. Images based on DR4 are slightly different than those of DR3, because of improvements made in the photometry pipeline. The $r$-band shapes of DR4 and DR3 images are not exactly the same, however, the alignment measurements from the two releases are quantitatively the same.

\section{Methodology}
\label{sec:Methodology}

\subsection{Radial alignment measurement}
\label{sec:radialalignment}

Measurements of alignments are often obtained by a statistical averaging of galaxy ellipticities. The ellipticity is defined in a Cartesian coordinate system as a non-linear manipulation of the galaxy surface brightness moments. The moments are given by 
\begin{equation}
Q_{ij}=\int G(\mathbf{x})\, x_1^i\, x_2^j\, \rm{d}\mathbf{x}\,,
\label{eq:unweighted moments}
\end{equation}
with $x_1,x_2$ the Cartesian coordinates where the galaxy is position at the coordinate system's origin and $G$ being the galaxy's surface brightness. The ellipticity is then computed from
\begin{equation}
\epsilon=\frac{Q_{20}-Q_{02}+{\rm i}2Q_{11}}{Q_{20}+Q_{02}+2\sqrt{Q_{20}Q_{02}-Q_{11}^2}}\,,
\label{eq:ellipticity}
\end{equation}
where $\epsilon = \epsilon_1+{\rm i} \epsilon_2$, and the ellipticity modulus is $|\epsilon|=(1-q)/(1+q)$, with $q$ being the semi-minor to semi-major axis ratio of the ellipse. This ellipticity measure is also often called the third flattening.

\begin{figure}
	\resizebox{\hsize}{!}{\includegraphics{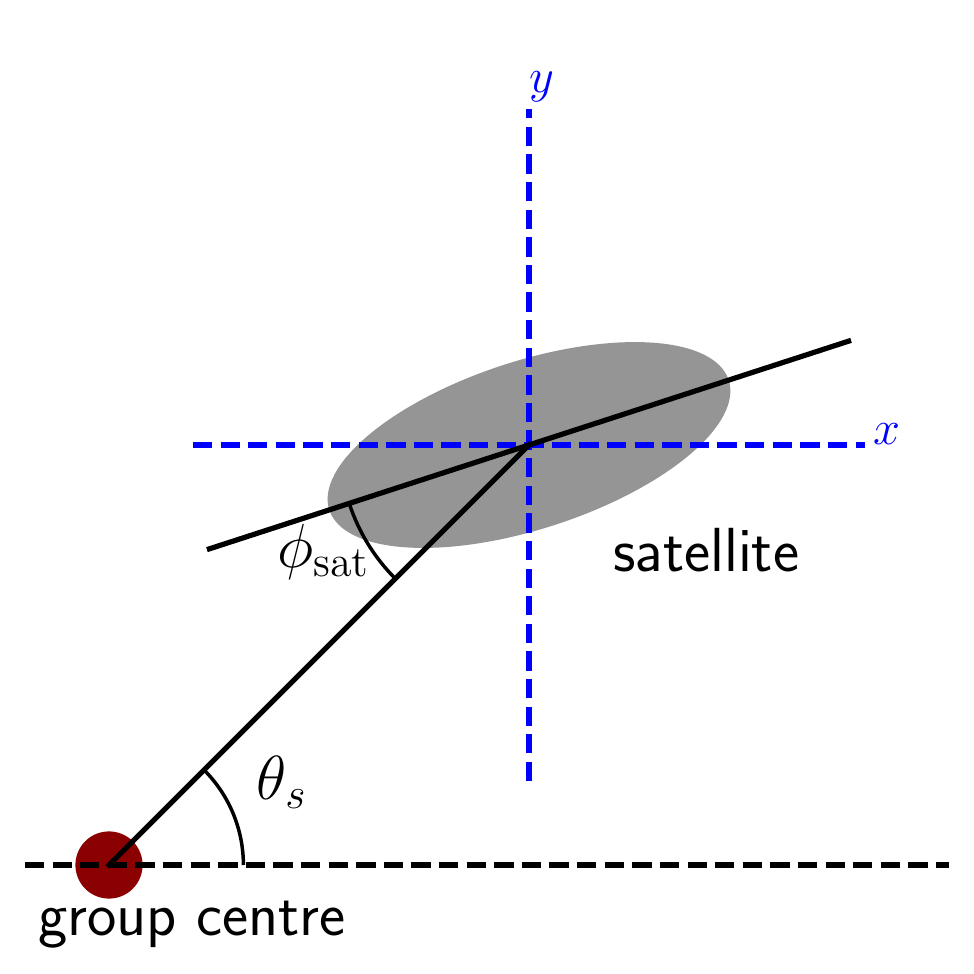}}
	\caption{Definition of the azimuthal angle of a satellite galaxy $\theta_{\rm s}$, as well as the angle between the line connecting the BGG and the satellite and the semi-major axis of the satellite $\phi_{\rm sat}$. The dashed black line coincides with the $x$-axis of the coordinate system, which is the same for every measured galaxy.}
	\label{fig:angles}
\end{figure}

To quantify satellite alignments, it is useful to rotate the Cartesian coordinate system so that the $x$-axis coincides with the vector pointing from the centre of the galaxy group to the centre of the satellite galaxy. In this frame, $\epsilon_1\mapsto\epsilon_+$ and $\epsilon_2\mapsto\epsilon_\times$, which are called tangential and cross ellipticity components, respectively. These are computed by 
\begin{align}
&\epsilon_+=\epsilon_1\cos2\theta_s+\epsilon_2\sin2\theta_s
\label{eq:eplus}\\
&\epsilon_\times=\epsilon_1\sin2\theta_s-\epsilon_2\cos2\theta_s\,,
\label{eq:ecross}
\end{align}
where $\theta_{\rm s}$ is the azimuthal angle of the satellite galaxy with respect to the group's centre (Figure \ref{fig:angles}).

When the tangential ellipticity is positive the satellite's semi-major axis points towards the centre of the group and the satellite is radially aligned with the group's centre. On the other hand, a negative value of $\epsilon_+$ means the satellite is tangentially aligned. Computing the average tangential ellipticity of a population of satellites will therefore reveal whether they preferentially align radially to the group's centre (in the case of a positive average $\epsilon_+$), tangentially (when negative) or their orientation is random (when the mean tangential ellipticity is zero). Note that the sign convention used in this work is opposite from the tangential ellipticity used in weak lensing analysis.

\subsection{Varying weight function}
\label{sec:weightfunction}

Measuring un-weighted moments, such as in Eq. \eqref{eq:unweighted moments}, from real astronomical images is not possible in practice. The most important reason for this is the presence of noise in the imaging data, whose contribution to the moments diverges as we go away from the galaxy's centre. One way to deal with this is to radially weight the galaxy's surface brightness, suppressing the noise at large separations. However, since the calculated moments are now weighted by this function, the resulting shapes will be biased. To minimise this, in the context of DEIMOS, a Taylor expansion is used to approximate the un-weighted moments by calculating higher order weighted moments \citep[see][for details]{DEIMOS}. We truncate the Taylor series at $n_{\rm w}=4$, which has been shown to be a good approximation and results in a smooth behaviour of biases associated with the overall shape measurement \citep{DEIMOS, Georgiou}. 

The weight function ideally needs to match the galaxy's profile, and a common choice is a 2D Gaussian function, which is an adequate approximation. Here, we use an elliptical 2D Gaussian function, whose ellipticity and centroid are matched iteratively per galaxy. The size of the Gaussian is preserved at each matching iteration. At first, the ellipticity of the Gaussian function is 0 and its initial scale $r_{\rm wf}$ is a free parameter, determined separately for each galaxy. This parameter will determine the galaxy scale taken into account during the shape measurement. A small initial $r_{\rm wf}$ will generally measure the shape of the inner regions of the galaxy (e.g. the bulge) while a large $r_{\rm wf}$ will reveal the shape of the outer galaxy regions (e.g. the disk).

We choose to relate the initial weight function scale $r_{\rm wf}$ to the galaxy's isophote radius $r_{\rm iso}$ at 3$\sigma$ above the background noise RMS, which is calculated using \textsc{SExtractor} \citep{sextractor}. We aim to probe the galaxy scale dependence of the intrinsic alignments by measuring the signal using shape measurements obtained for several values of $r_{\rm wf}$. We use a total of 5 different weight functions, namely $r_{\rm wf}/r_{\rm iso}=\{0.5, 1, 1.25, 1.5, 2\}$, and restrict this to $r$-band images, which are higher quality, taken under better observing conditions compared to $g$ and $i$. On shapes from $g$ and $i$-band observations, the weight function size is fixed $r_{\rm wf}=r_{\rm iso}$, where the isophote is measured in the deepest $r$-band images. 

The physical regions probed by a certain weight function vary for each galaxy, depending on the physical size probed by the  isophote of the galaxy. The mean isophote size in our galaxy sample is around 10 kpc with a standard deviation of 3.7 kpc. Therefore, even for the largest weight function we considered, the physical regions probed for each galaxy do not extend beyond the galaxy itself.

\subsection{Tests for systematic errors}
\label{sec:systematics}

Here we discuss a few sources of potential contamination to our satellite alignment signal measurement and several ways to quantify and control this contamination, in order to robustly measure the alignment of satellite galaxies.

\subsubsection{Shape measurement bias}
\label{sec:bias}

There are several effects that can produce a systematic error when measuring galaxy shapes. This is often characterized with the multiplicative $m$ and additive $c$ bias, through the equation \citep{Heymans},
\begin{equation}
\epsilon_i^{\rm{obs}}=(1+m_i)\epsilon_i^{\rm{true}}+c_i\,,
\label{eq:bias}
\end{equation}
where the observed galaxy ellipticity is related to the true ellipticity and the index runs through the two ellipticity components. An additive bias in the shape measurement method (often a product of incorrect treatment of the PSF) can bias the alignment measurement. However, as we compute the average tangential ellipticity component, such additive biases will average to zero. Indeed we show in Section \ref{sec:BCGcontamination} that we do not measure an additive bias caused by the shape measurement procedure alone in our image simulations. However, an additive $c_+$ term caused by the light of the BGG can offset our alignment signal \citep[see e.g.][for a similar analysis]{Sifon18}, and we quantify this in Sect. \ref{sec:BCGcontamination}.

Another bias that affects the alignment signal is multiplicative bias (or $m$-bias). This is unavoidable when measuring shapes from noisy images, due to the non-linear manipulation of pixel data, required to obtain shape information \citep[see e.g.][]{Viola2014}. Fortunately, the $m$-bias can be determined through image simulations of galaxies that represent the overall galaxy sample. It is important to construct realistic and representative simulations \citep{Hoekstraimsim} as well as to use a shape measurement method that is robust, with shape measurement accuracy that depends weakly on galaxy properties. In \citet{Georgiou} it was shown that DEIMOS applied on simulations of GAMA galaxies, which have generally high signal-to-noise ratio (SNR) in KiDS images, does not depend strongly on input ellipticity, size or SNR.

Following the work in \citet{Georgiou}, we simulate galaxy images using the \textsc{GalSim} python package \citep{Galsim}. We obtain galaxy morphological parameters from the GAMA Sersic Photometry Data Management Unit (DMU) which includes single component S\'ersic profile fits to reprocessed $r$-band Sloan Digital Sky Survey images \citep{Kelvin}. We perform certain quality control cuts that are present in the catalogue ({\tt GAL\_QFLAG}, {\tt GAL\_GHFLAG} and {\tt GAL\_CHFLAG} equal to zero, which ensures there were no problems during the parametric fit). We also avoid simulating galaxies with S\'ersic indices $n_{\rm s}<0.3$ and $n_{\rm s}>6.2$ due to limitations of the \textsc{GalSim} \citep{Galsim}. In these cases we set $n_{\rm s}=0.3$ or $6.2$, respectively, which, as discussed in \citet{Georgiou}, does not affect the bias calibration significantly. Galaxy profiles are truncated at $4.5$ times their half-light radius. The input flux and ellipticity of the galaxy image simulations are measured from the real KiDS images. Galaxies are simulated on a grid, with a $300\times300$ pixels postage stamp. The simulated PSF is an elliptical Gaussian with ellipticity and full width at half maximum matching the median value of real KiDS images. Finally, we take into account the effect of masking neighbouring galaxies by using the segmentation map of each simulated galaxy as measured in the KiDS images from \textsc{SExtractor}. The masked pixels are replaced with Gaussian noise of the same RMS as the image, to create an individual postage stamp for each galaxy from which the shape is measured.

Using these image simulations we calculate the bias for the five different weight functions considered in this work, with simulations of KiDS $r$-band images. We confirm that values of $m_1$ and $m_2$ agree within the statistical error bars of their measurement and, for simplicity, we use the average value $\langle m\rangle=(m_1+m_2)/2$. The $m$-bias was found to be relatively small and is corrected for when computing the average ellipticity components. The values are presented in Table \ref{tab:bias}.

\begin{table}
	\caption{Mean multiplicative and bias obtained from image simulations for the various weight functions used in the $r$-band shape measurement.}
	\label{tab:bias}
	\centering
	\begin{tabular}{l c }
		\hline\hline
		$r_{\rm wf}/r_{\rm iso}$ & $\langle m \rangle$ \\
		\hline 
		0.5 & $0.02145\pm0.00089$ \\
		1.0 & $-0.00373\pm0.00084$ \\
		1.25 & $-0.00652\pm0.00121$ \\
		1.5 & $-0.01120\pm0.00181$ \\
		2.0 & $-0.04945\pm0.00397$ \\
		\hline
	\end{tabular}
\end{table}

We also look into the radial satellite alignment at different broad band filters and for these we obtain bias values from \citet{Georgiou} (note that when comparing the different filters we use the same weight function $r_{\rm wf}=r_{\rm iso}$). While the galaxy population considered in this work is not the same as in our previous study (i.e. here we only consider shapes of group galaxies and not of the full GAMA galaxy sample) and the $m$-bias for it could be different, it has been shown in \citet{Georgiou} that the bias does not depend on galaxy properties very strongly for the GAMA galaxy sample. Therefore we choose to use this previously determined $g,r$ and $i$-band shape $m$-bias estimation for our calibration.

Our sample of galaxies has a high SNR (mean value of 300 for $r$-band images) and the shape measurements are robust against galaxy properties. For example, shapes depend very weakly on galaxy ellipticity \citep{Georgiou}. This guarantees that any trend observed of the alignment signal for different galaxy sub-samples is robust against ellipticity measurements.

\subsubsection{``Stray'' BGG light}
\label{sec:BCGcontamination}

An important systematic effect that could bias the satellite alignment signal is the contamination from light of the group's brightest galaxy, the BGG. The light measured in images of satellite galaxies includes the light profile of the satellite as well as the faint ``tail'' from the light profile of the BGG. Ultimately, this will lead to the background light of the satellite being slightly ``tilted'' towards the direction of the group's centre, meaning the side of the satellite closer to the group's centre will have a slightly higher background compared to the side further away from it \citep[see also][]{Sifon18}. Since to extract the alignment signal we average the tangential ellipticities of all satellite galaxies, this effect gets amplified and can increase the radial alignment signal or even produce a completely artificial one. The BGG contamination will be more severe for satellites closer (in projection) to the group's centre, as well as for shapes measured using a higher weight function, which will allow more of the BGG's light to impact the shape measurement.

In order to quantify and correct for the contamination from the BGG's ``stray'' light we again use image simulations, very similarly to \ref{sec:bias}. We emulate the GAMA galaxy groups by simulating galaxies for each group in their relative positions. The satellite galaxy profiles are truncated at a distance of 4.5 times their half-light radius while for the BGGs, simulated in the centre of the image, the light extends throughout the whole image. The input ellipticity of each satellite is such that the ellipticity modulus, $|\epsilon|$, is obtained using DEIMOS on KiDS $r$-band images with a weight function $r_{\rm wf}=r_{\rm iso}$ and $\epsilon_1,\epsilon_2$ are constructed using a random position angle. This ensures that the input average ellipticity is zero, i.e. the satellites are randomly oriented in our image simulations. The input ellipticity of the BGG is the same as the one observed in the real images. Even though the input ellipticity is based on one weight function, the output ellipticity is corrected for $m$-bias, which is anyway negligibly small, and our results are not affected by this.

\begin{figure}
	\resizebox{\hsize}{!}{\includegraphics{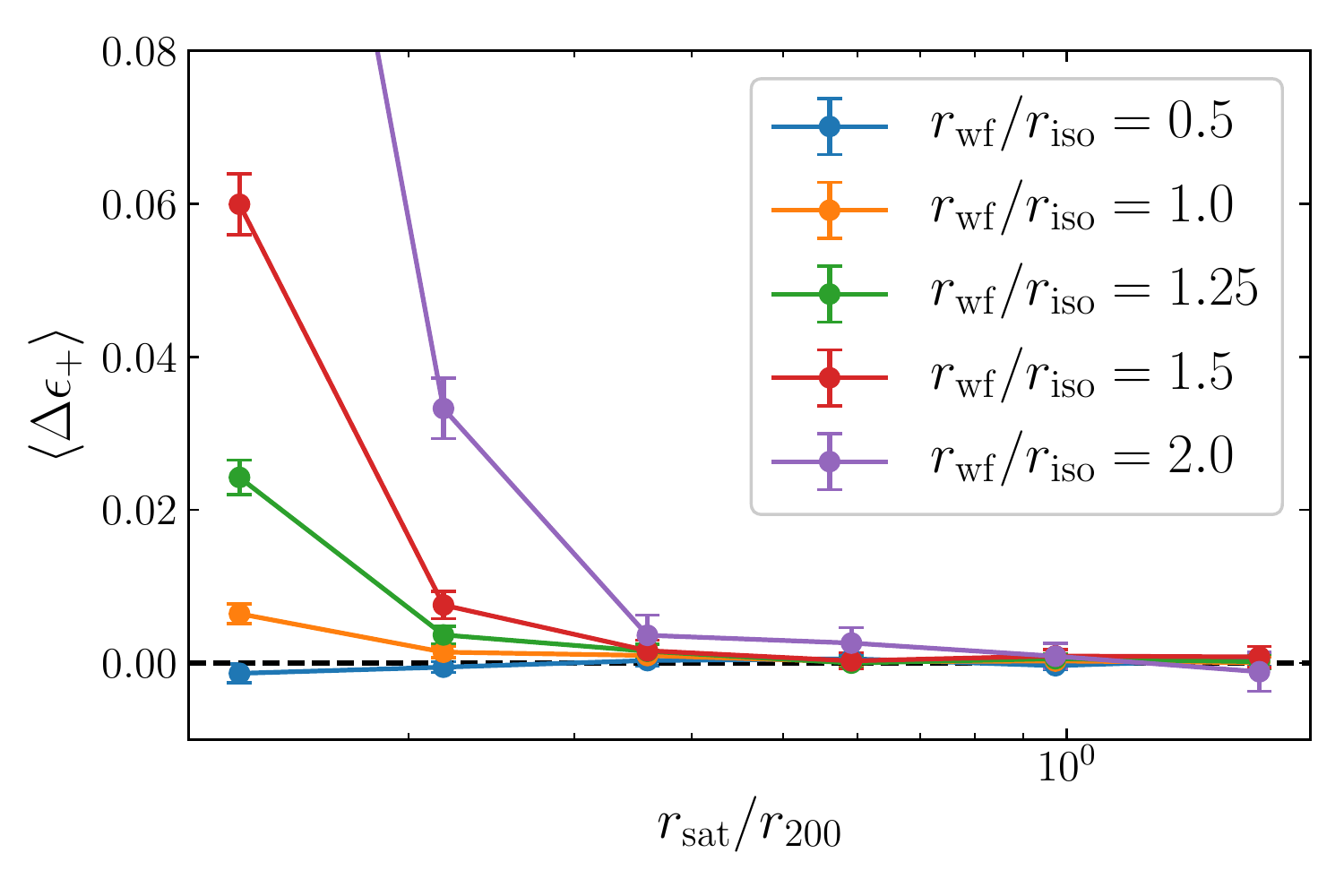}}
	\caption{Mean tangential ellipticity as a function of distance from the group's center, calculated using $\epsilon^{\rm out}-\epsilon^{\rm in}$ from simulated KiDS $r$-band images of GAMA groups. The input ellipticity is constructed using the measured $|\epsilon|$ for each satellite galaxy and assigning a random position angle. The output ellipticity is measured from the images, which all contain the corresponding BGG in the centre. Therefore the mean tangential ellipticity difference shows the contribution of the BGG's stray light to the satellite's radial alignment signal measurement.}
	\label{fig:BCGcontamination}
\end{figure}

We measure the shapes of the satellite galaxies in the simulated $r$-band images, which are affected by the light from the BGG. We then subtract the input ellipticity from the output and compute the average tangential ellipticity difference $\langle\Delta\epsilon_+\rangle$. The results are presented if Figure \ref{fig:BCGcontamination} with the $y$-axis showing the contamination from the BGG in the alignment signal and the $x$-axis the projected physical distance of the satellite from the BGG, normalized by the group's $r_{200}$ (see Sect. \ref{sec:Results}). As expected, satellites closer to the BGG and shapes with a larger weight function are more heavily contaminated by the BGG's light. 

We note that, when comparing the alignment signal between different filters, we use the same correction for the BGG's light. We check that quantitatively similar results are obtained with $g$ and $i$-band image simulations, with larger error bars, hence we use $r$-band simulations also for calibrating the alignment measured in the other two bands. 

The correction of the BGG's light to the alignment signal is heavily influenced by the fidelity of our image simulations. While simulating galaxies that are representative of the galaxy sample is essential, it is also important to understand the actual light profiles of these galaxies. In our image simulations we assume galaxies can be represented by a single S\'ersic profile, but this is a simplification since in reality galaxies have complicated morphologies, such as bulges, bars and star-forming regions. In addition, the isophotes of the BGGs twist their position angles the further away they are, which is not included in our simulations. Also, we choose to truncate the satellite's profile and leave the BGG S\'ersic profile to extend indefinitely. Finally, we do not capture the effect of background subtraction applied in the reduction of the KiDS imaging data, which can affect the faint tail of the BGG light. Doing so would require a full reduction of image simulations in a similar manner to KiDS images but would not eliminate all the uncertainties mentioned, and we choose not to do so. Hence, we cannot guarantee that the correction applied is accurate enough. However, as long as the correction is small, compared to a non-zero intrinsic alignment signal measurement, we can be confident that the measurement is robust against the contamination from the BGG's light.

Lastly, we produced a set of image simulations with the satellite galaxies but without including the BGG. The input alignment signal of these simulated satellites is also zero. Therefore, any bias caused solely by the shape measurement method would be detected in these simulations. Such bias is not found, as the alignment signal measured from the output ellipticities was also measured to be zero.

\section{Satellite galaxy alignments}
\label{sec:Results}

We present the satellite radial alignment signal as obtained through measurements of the average tangential ellipticity of our satellite galaxy sample, with respect to the position of the group's BGG. In all of our measurements we correct for multiplicative bias (Table \ref{tab:bias}) as well as subtracting the artificial signal from the BGG's light (Figure \ref{fig:BCGcontamination}). The shape measurement method is capable of flagging measurements of ellipticity that were problematic (e.g. when the centroid determination failed or the image moments are nonsensical, see also \citealt{Georgiou}). We reject such galaxies and only consider objects with well defined ellipticity. We calculate the virial mass and radius for each group using the group's luminosity (column \texttt{LumB} in the galaxy group DMU). This is an unbiased measure of the group's luminosity, calibrated against specialized mock galaxy catalogues \citep[see][for details]{Aaron}. Using the scaling relation computed specifically for GAMA galaxy groups in \citet{Violaetal}, we get the virial mass from
\begin{equation}
\left(\frac{M_{200}}{10^{14} M_\odot h^{-1}}\right) = (0.95\pm0.14)\left(\frac{L_{\rm grp}}{10^{11.5}L_\odot h^{-2}}\right)^{1.16\pm0.13}\,,
\label{eq:scaling_relation}
\end{equation}
where $L_{\rm grp}$ is the group luminosity, and the virial radius from
\begin{equation}
M_{200}=\frac{4\pi}{3}\rho_{\rm c} r_{200}^3\,,
\end{equation}
with $\rho_{\rm c}$ the critical density of the universe. We look into the dependence of the signal on the projected distance of the satellites from the group's centre $r_{\rm sat}$, normalized by the group's $r_{200}$. Trends of the alignment signal with satellite absolute magnitude (but not with the BGG's magnitude) have been observed in clusters \citep{Huang2}, and we examine this in groups. Groups are also binned in mass, since groups with different gravitational potential could exhibit differences in their alignments. To identify possible trends, we fit the satellite alignment measurements with a power-law function of the form $\langle\epsilon_+\rangle=A(r_{\rm sat}/r_{200})^g$, where $A$ and $g$ are the amplitude and index of the power-law. Fitting for both these parameters yields very weak constraints, and we choose to fix one, specifically the power-law index, to the value of $g=-2$, which seems to describe the data adequately. This arbitrary choice of fitting function is justified by the fact that we are not trying to find a physical model for the alignment signal but rather quantify trends and significance, hence a fully empirical description is enough. We fit our data using the non-linear least squares fitting method from the \textsc{scipy} python package. The error bars in our measurements are standard errors. We have checked that the
mean tangential ellipticity distribution is Gaussian and the standard error
should accurately reflect the significance of a detection. We use groups with 3 or more members (including the BGG). Our results do not change within our error bars when using groups with 5 or more members.

\subsection{Full sample and wavelength dependence}
\label{sec:colordependence}

\begin{figure}
	\resizebox{\hsize}{!}{\includegraphics{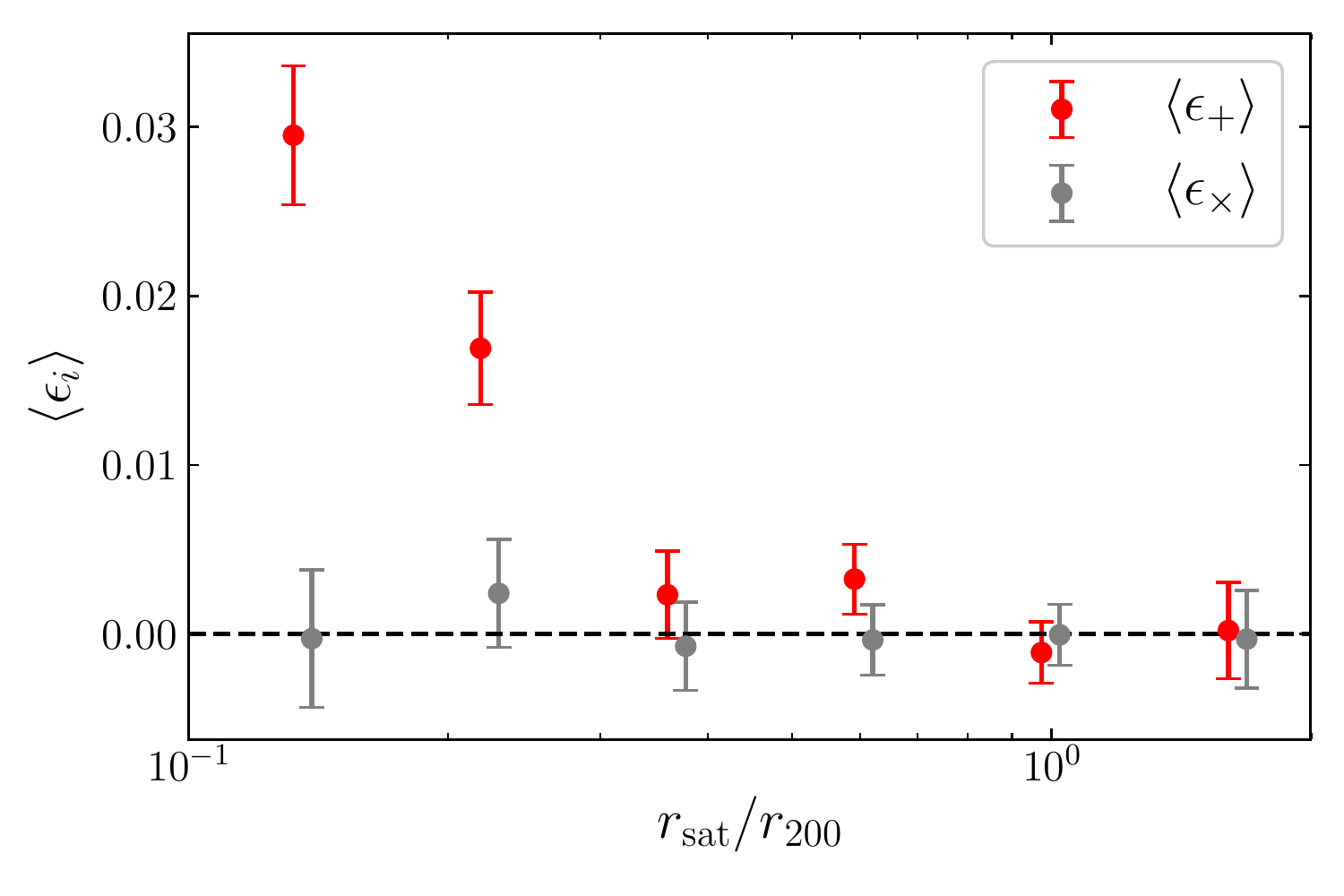}}
	\caption{Mean tangential/cross ellipticity (red/gray points) versus projected satellite distance from the group's BGG, obtained using satellite shape - BGG position pairs. Positive values of $\langle\epsilon_+\rangle$ indicate radial alignment while randomly oriented galaxies exhibit $\langle\epsilon_+\rangle=0$. Results shown for shape measurements obtained from $r$-band images, with $r_{\rm wf}=r_{\rm iso}$.}
	\label{fig:ep_rband}
\end{figure}

We measure the satellite radial alignment signal for our galaxy group sample. We reject failed shape measurements as well as galaxies residing in masked regions (except for secondary and tertiary stellar halos, which are expected to be too dim to affect the shape measurements). This leaves us with 29953 satellite-BGG pairs. We first present results for shapes measured in $r$-band images, since these are the deepest observations with the smallest shape noise in our sample. The weight function used is $r_{\rm wf}=r_{\rm iso}$, and we use this weight function for all subsequent measurements, unless otherwise stated. The results are shown in Figure \ref{fig:ep_rband}. Significant radial alignments are observed for satellites close to the BGG. The alignment signal decreases for satellites further away, eventually reaching zero at scales $r_{\rm sat}\sim r_{200}$. We also note that the cross ellipticity component is consistent with zero, which holds true for all subsequent measurements.

\begin{figure}
	\resizebox{\hsize}{!}{\includegraphics{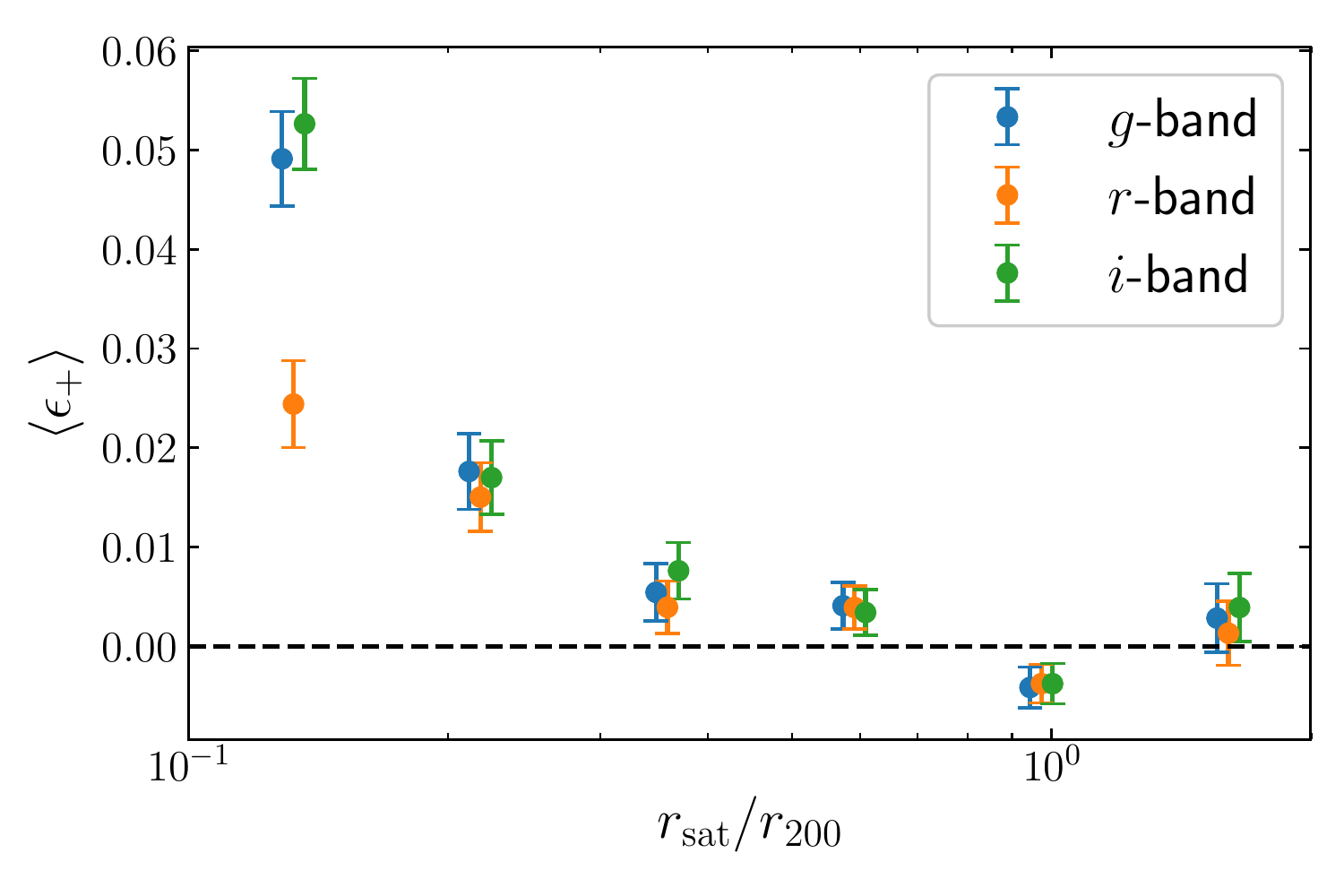}}
	\caption{Mean tangential ellipticity versus satellite distance from the group's BGG, obtained using satellite shape - BGG position pairs. Positive values indicate radial alignment, randomly oriented galaxies give $\epsilon_+=0$. Results shown for shape measurements obtained from $g,r$ and $i$-band images.}
	\label{fig:ep_allband}
\end{figure}

We then compute the alignment signal using shapes from \citet{Georgiou} (which are based on KiDS DR3, see Sect. \ref{sec:shapes}), for $g$, $r$ and $i$-band observations. We consider only galaxies that are not masked or flagged in any of the three filters, and compare the alignment signals in Figure \ref{fig:ep_allband}. It is clear that in the smallest radial bin $r_{\rm sat}\sim0.13r_{200}$ the alignments in $g$ and $i$ are much larger than the $r$-band. Even though the BGG's ``stray light'' correction for this radial bin is large ($\sim$20\% of the $r$-band signal in Fig. \ref{fig:ep_rband}) and robust quantitative conclusions are hard to draw, there is a significant excess alignment signal in the first bin, observed in $g$ and $i$-band in comparison to the $r$-band. A possible concern is that bright sources (such as the brightest group galaxy) have a higher flux in $r$-band compared to $g$- or $i$-band, and therefore the background subtraction of the area around this bright source can be more aggressive in $r$-band images compared to $g$- and $i$-band. However, we do not look into this further in this work.

\subsection{Absolute magnitude dependence}
\label{sec:magnitudedependence}

\begin{figure}
	\resizebox{\hsize}{!}{\includegraphics{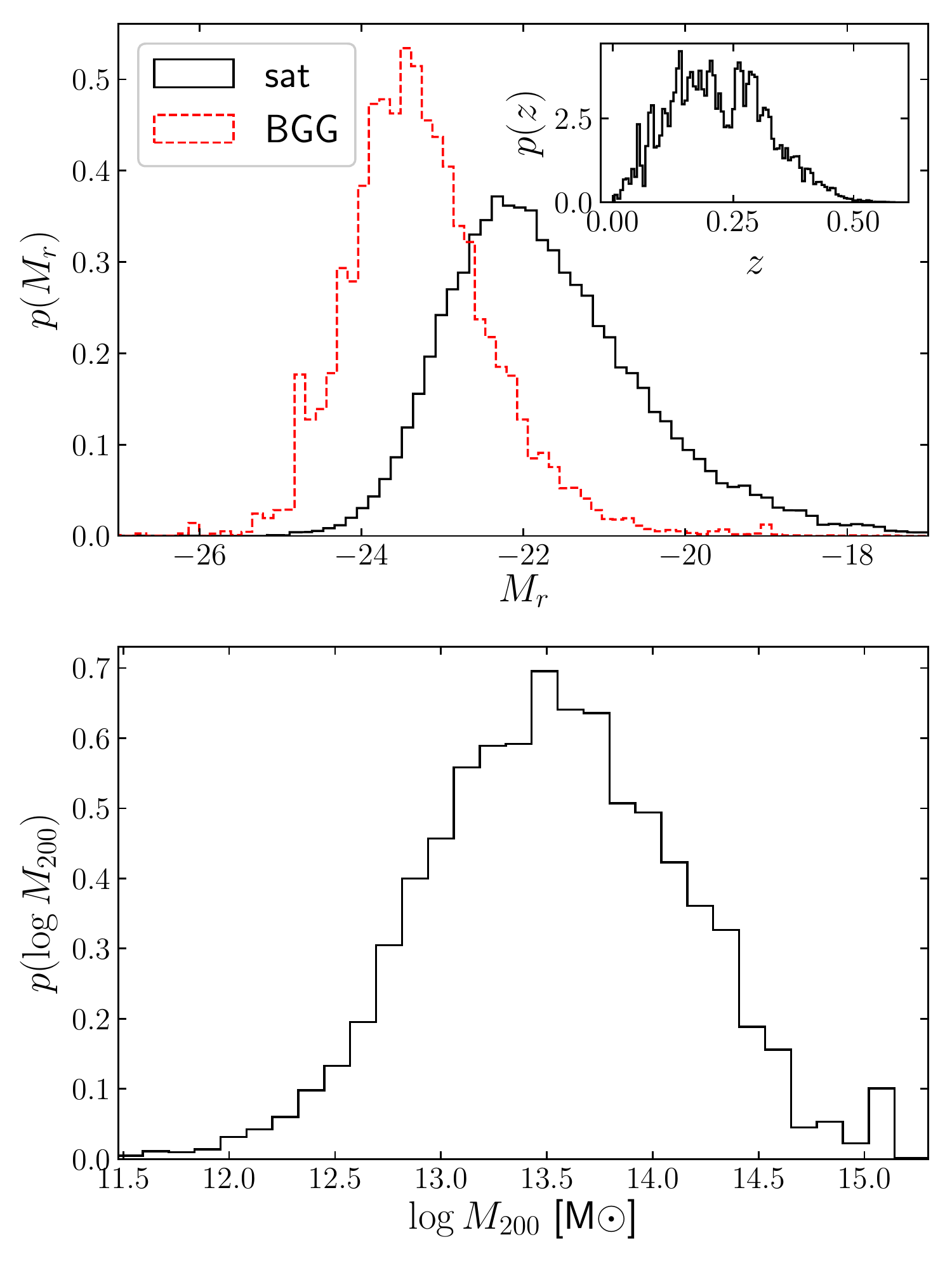}}
	\caption{\emph{Top}: Histograms of the absolute r-band magnitude of the BGG (solid black) and satellite (dashed red) galaxy samples. The inset figure shows the distribution of the galaxy group redshifts. \emph{Bottom}: Histogram of the group virial mass, computed with Eq. \eqref{eq:scaling_relation}.}
	\label{fig:MrM200hist}
\end{figure}

\begin{figure}
	\resizebox{\hsize}{!}{\includegraphics{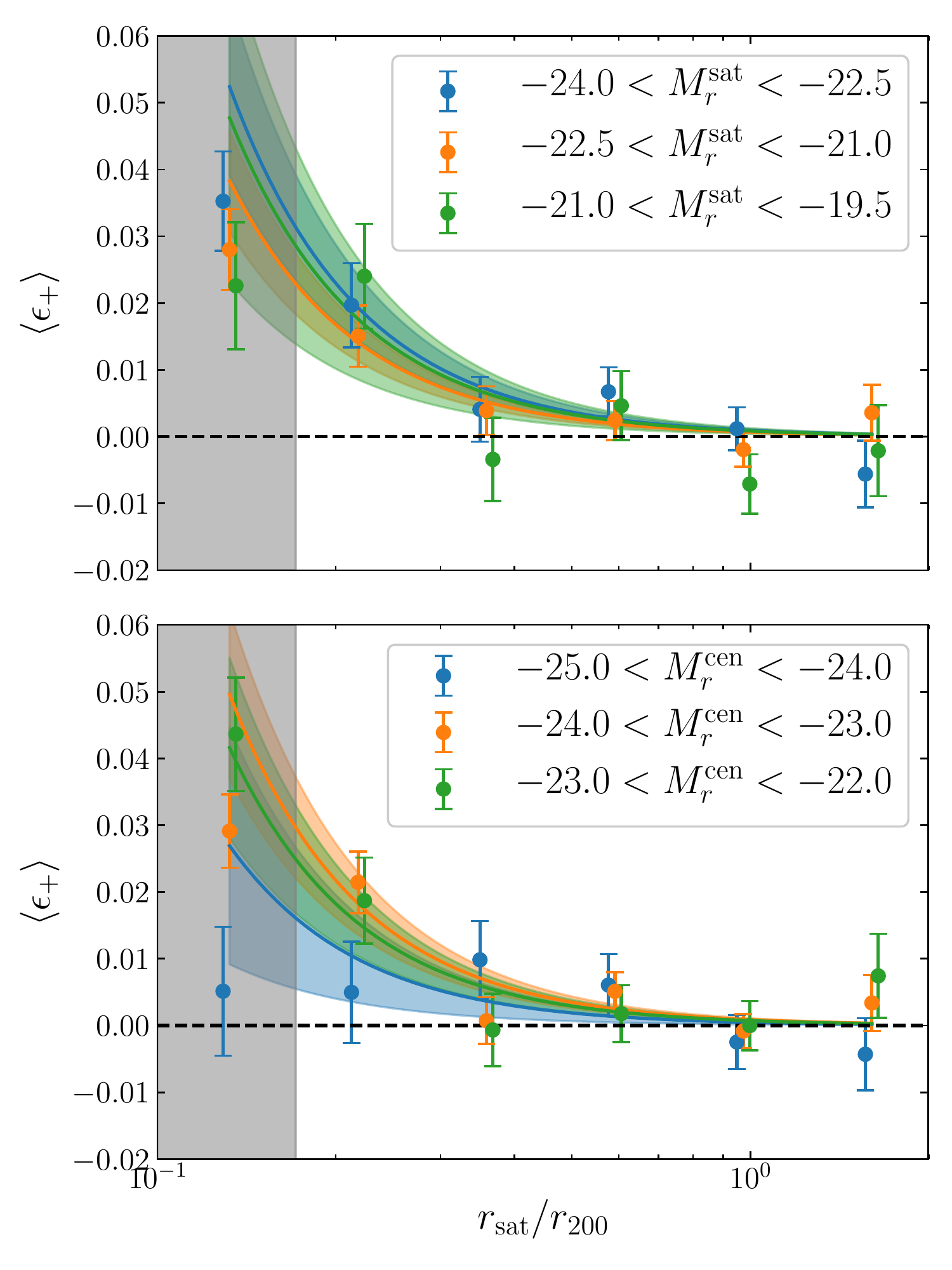}}
	\caption{Mean tangential/cross ellipticity components versus satellite distance from the BGG, obtained using satellite shape - BGG position pairs. \emph{Top:} binned in satellite's $r$-band absolute magnitude. \emph{Bottom:} binned in BGG's $r$-band absolute magnitude. Power-law fits with fixed index are over-plotted, with the 1-$\sigma$ uncertainty on the fit, for the three bins. The grey area covers the radial bins over which we regard the BGG light correction to be too significant, and these scales are not used for our model fitting. Shapes were measured in $r$-band for $r_{\rm wf}=r_{\rm iso}$.}
	\label{fig:ep_Mr}
\end{figure}

We study the dependence of satellite alignments on the r-band absolute magnitude of satellites and centrals. Magnitudes are obtained from stellar population fits to galaxy SEDs in multiple bands using \textsc{Lambdar}, made available through the GAMA \texttt{StellarMassLambdarv20} DMU \citep{StellarMasses, Lambdar}. Histograms can be seen in the top panel of Fig. \ref{fig:MrM200hist}. We first look at the dependence on the satellite magnitude, in Figure \ref{fig:ep_Mr}, top panel. In order to quantify any dependence on satellite magnitude, we fit a power-law with a fixed slope of -2 to all radial bins except the first (due to the BGG light correction being too large to make robust quantitative statements about this bin). The fit is also shown in Figure \ref{fig:ep_Mr}, with the 1-$\sigma$ confidence intervals. No clear trend is observed for the three magnitude bins within the 1-$\sigma$ limit.

Similar results are obtained when considering the central galaxy's absolute magnitude. This is shown in Figure \ref{fig:ep_Mr}, bottom panel, where again the 1-$\sigma$ limits are not strong enough to identify any trends in the alignment signal. When looking at galaxy clusters, \citet{Huang2} found a dependence of the alignment with satellite absolute magnitude, which we do not detect. This could mean that this dependence is stronger in clusters than groups. However, the galaxy shapes in \citet{Huang2} were not corrected for shape measurement biases. Multiplicative bias scales with the galaxy SNR, which can be correlated with its absolute magnitude, and can drive part or all of this dependency, so a robust conclusion is unclear.

\subsection{Group mass dependence}
\label{sec:massdependence}

\begin{figure}
	\resizebox{\hsize}{!}{\includegraphics{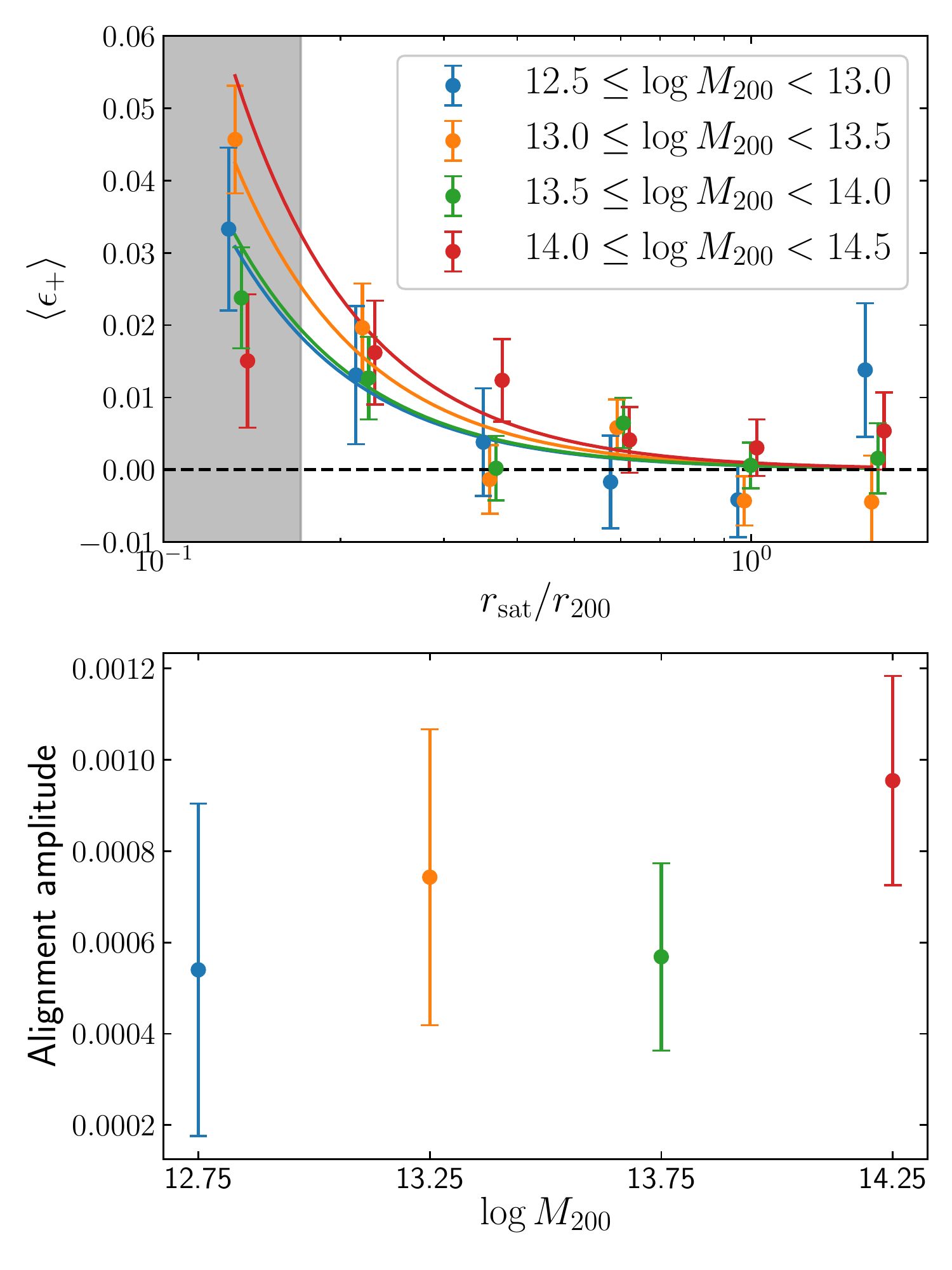}}
	\caption{\emph{Top:} mean tangential/cross ellipticity as a function of satellite distance from the group's BGG, obtained using satellite shape - BGG position pairs, binned in the group's $M_{200}$. The best fit power-law with a fixed slope is overplotted for each mass bin. The grey area covers the radial bins over which we regard the BGG light correction to be too significant, and these scales are not used for our model fitting. Shapes were measured in $r$-band for $r_{\rm wf}=r_{\rm iso}$. \emph{Bottom:} Amplitude of the power-law fit to the alignment signal for each individual group mass bin.}
	\label{fig:epM200}
\end{figure}

Here we look into the dependence of the satellite alignment signal for groups of different mass. We calculate virial masses of groups according to equation \eqref{eq:scaling_relation}. A histogram of the resulting group masses is shown in the bottom panel of Fig. \ref{fig:MrM200hist}. We present the average tangential ellipticity versus distance of satellite to the group's BGG, binned in $M_{200}$, in Figure \ref{fig:epM200}. We fit a power-law with fixed amplitude as previously, but do not see any significant trend within the 1-$\sigma$ limit.

\subsection{Dependence on star formation rate}

We now distinguish our satellite galaxy sample to satellites that may have arrived in the group recently or have been orbiting for a longer period. Galaxies entering a group will experience star formation quenching through several processes, such as tidal stripping. Therefore, these members will have mostly old stellar populations and appear red in colour. Galaxies that have recently fallen into the group will generally have a higher star formation rate with new stars constantly generated, and hence will appear bluer (except for galaxies that are already red). Thus, a proxy for distinguishing between ``old'' and ``new'' group members would be using colour and specific star formation rate information.

\begin{figure}
	\resizebox{\hsize}{!}{\includegraphics{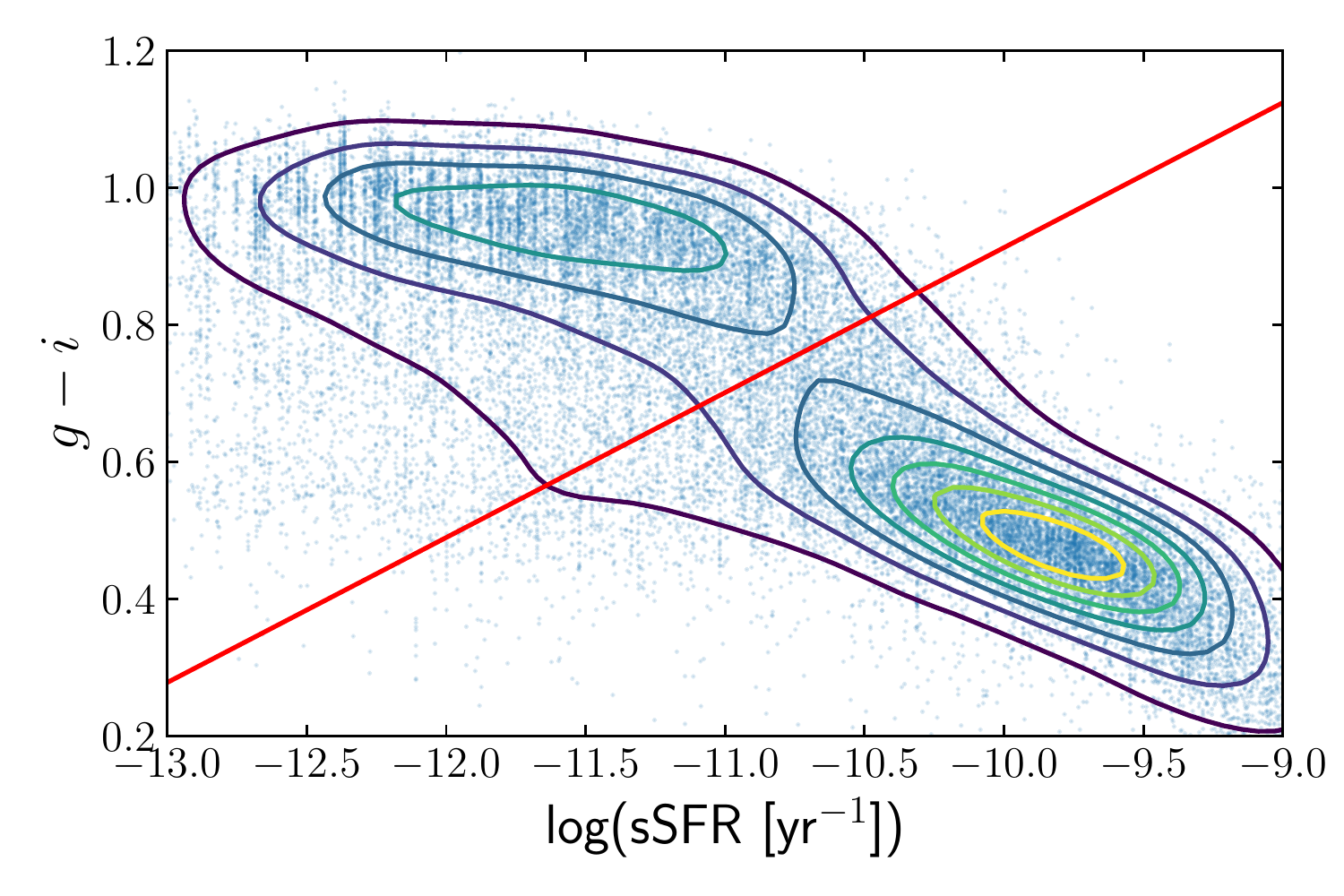}}
	\caption{Rest-frame, dust-corrected $g-i$ colour of our satellite galaxy sample against their specific star formation rate. Smoothed density contours are shown, with dense regions appearing more yellow. The red line indicates the cut performed to divide our galaxy sample into blue star forming, and red galaxies with lower star formation rate.}
	\label{fig:gmi_sSFR}
\end{figure}

We use the GAMA \texttt{StellarMassesLambdarv20} catalogue to obtain dust-corrected, rest-frame $g-i$ colour information for our satellite galaxy sample, and the \texttt{MagPhysv06} GAMA DMU \citep{GAMADR3, MagPhysDMU} which contains specific star formation rate (sSFR) measurements acquired from running the SED-fitting code \textsc{MagPhys} \citep{Magphys}. We show the $g-i$ colour of satellites against the sSFR in Figure \ref{fig:gmi_sSFR}. It is clear that the galaxies are divided into two distinct populations: blue galaxies with high star formation rate (in the bottom right of the Figure) and red galaxies with low star formation rate (in the top left of the Figure). By visual inspection, we divide our galaxy sample into blue and red galaxies, using the red line in Figure \ref{fig:gmi_sSFR}, 
\begin{equation}
g-i = 0.21 \log({\rm sSFR\, yr^{-1}}) + 3.03\,.
\label{eq:bluered}
\end{equation}

\begin{figure}
	\resizebox{\hsize}{!}{\includegraphics{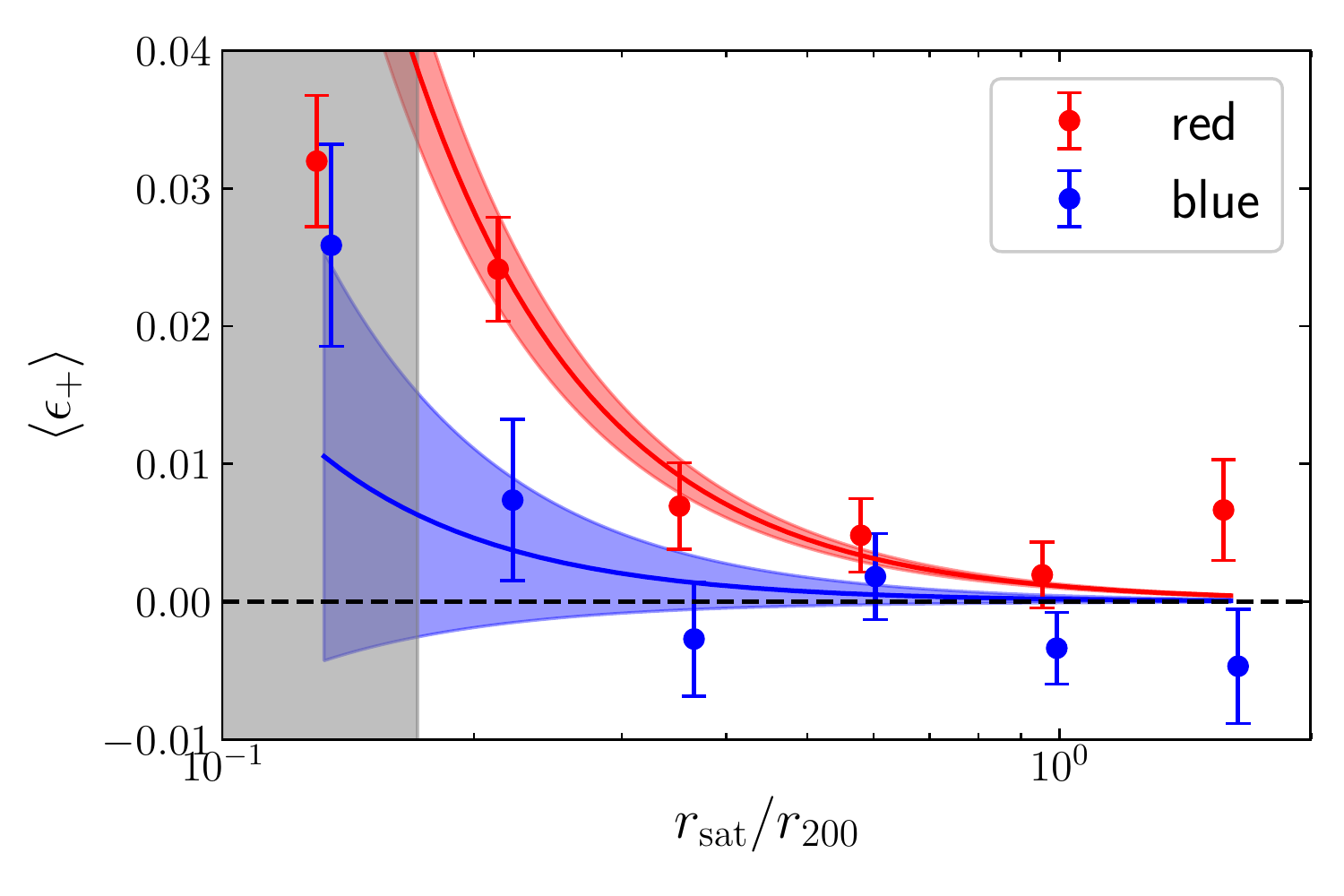}}
	\caption{Average tangential ellipticity against projected satellite distance from the group BGG, obtained using satellite shape - BGG position pairs. Results shown for red, low sSFR and blue, high sSFR galaxies. Power-law fits of fixed index are overplotted, with the shaded region indicating the 1-$\sigma$ confidence limit. The grey area covers the radial bins over which we regard the BGG light correction to be too significant, and these scales are not used for our model fitting.}
	\label{fig:ep_sSFR}
\end{figure}

We compute the alignment signals of red and blue galaxies and show the results in Figure \ref{fig:ep_sSFR}. It is clear that red satellites have a much stronger alignment signal than blue ones. This behaviour can be explained considering that red galaxies have likely spent more time in the group environment and have had more time to interact gravitationally with the group and become aligned with it compared to blue satellites that have likely been in the galaxy group for less time. Blue satellites are also generally rotationally supported and have a different alignment mechanism than red, pressure supported galaxies. We note that the difference in the alignment signal observed here cannot be attributed to differences in the ellipticity distributions of the two sub-samples, as the ellipticity estimation for our sample does not depend strongly on the galaxy ellipticity \citep{Georgiou}.

\subsection{Galaxy scale dependence}
\label{sec:scaledependence}

Lastly, we perform the satellite radial alignment measurement with shape measurements obtained using weight functions of different sizes. This results in shape measurements being more sensitive to the different parts of the galaxies. For example, a smaller weight function will mostly probe the shape of the inner part of a galaxy (e.g. the bulge) while a larger weight function will be more sensitive to the shape of the outer part (e.g. the disk). We measure shapes of galaxies using five different weight function sizes $r_{\rm wf}$, which corresponds to the scale of the initial (circular) Gaussian weight function applied when measuring galaxy shapes. This weight function eventually matches the ellipticity of the galaxy after several matching iterations, while retaining its original size (see Sect. \ref{sec:weightfunction}). 

\begin{figure}
	\resizebox{\hsize}{!}{\includegraphics{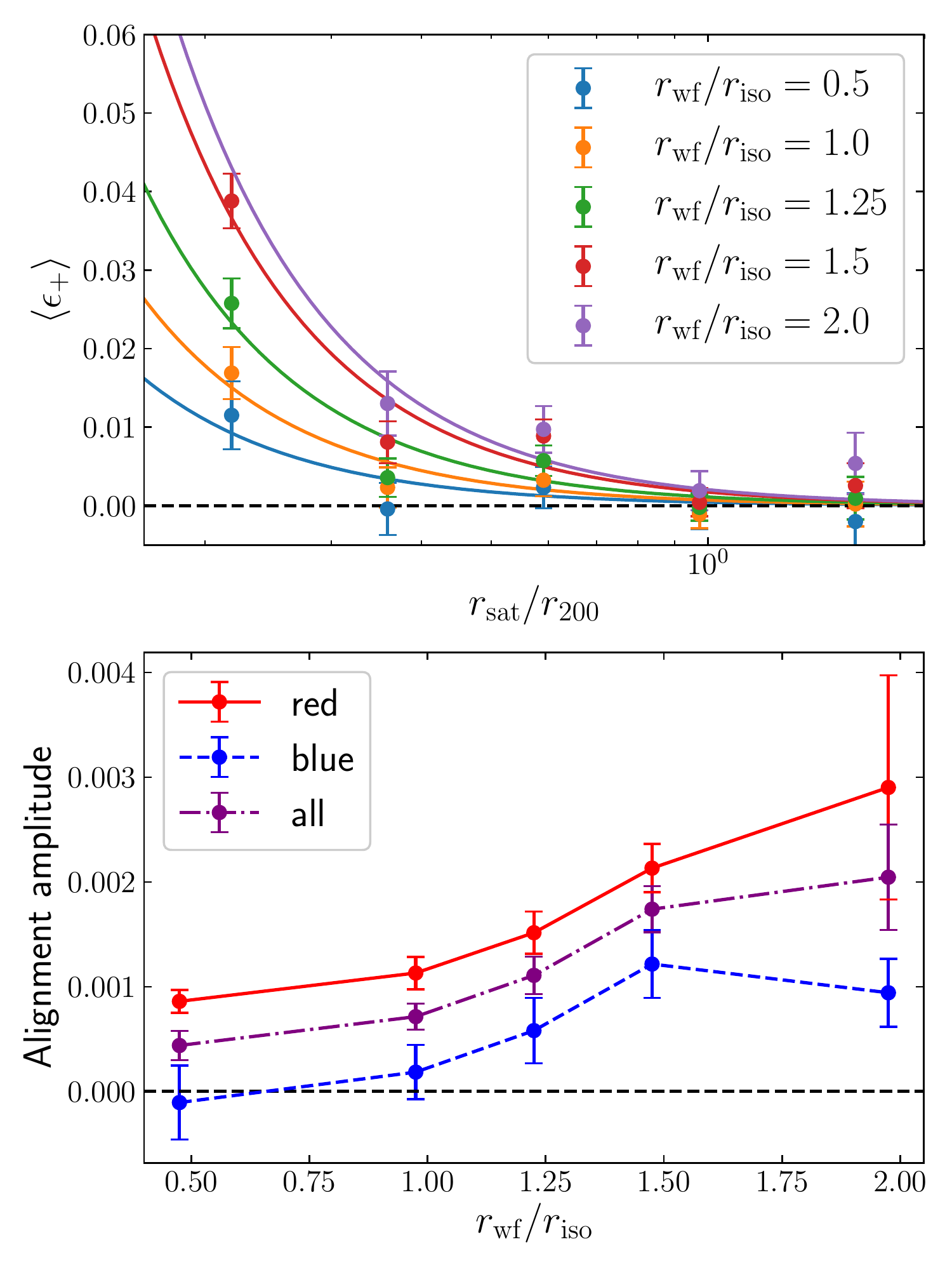}}
	\caption{\emph{Top:} Average tangential ellipticity of the \emph{full} satellite galaxy sample versus projected satellite distance from the group BGG, obtained using satellite shape - BGG position pairs. Results shown for different weight function sizes used in the shape measurement process. The $r_{\rm wf}$ indicates the size of the initial weight function and $r_{\rm iso}$ is the isophote of the galaxy. \emph{Bottom:} Amplitude of power-law fits with fixed exponent, to the average ellipticities for all, red and blue galaxies, for the different weight function sizes involved. Lines connecting the data points added for clarity.}
	\label{fig:ep_wf}
\end{figure}

In Figure \ref{fig:ep_wf}, we show the alignment signal for the full satellite sample, as computed for the five different weight function sizes that are related to the isophote at 3$\sigma$ above the background noise. Note that the $x$-axis of this plot has different limits than similar figures in the section. We use only 4 instead of 5 bins for the largest weight function, which suffers more severely from contamination from the BGG's light. Fitting a power-law model with a fixed index of $-2$ we can see that the alignment signal increases for larger weight functions (bottom panel of Figure \ref{fig:ep_wf}). In the bottom panel of the figure we also show the alignment amplitude for red and blue galaxies. We see that blue galaxies exhibit an increase in alignment for larger weight functions, but are always smaller than red galaxies. For the larger weight function sizes, the blue alignment signal is significantly non-zero.

We note here that the ellipticity distribution of galaxies obtained from the 5 different weight functions are not identical, as is expected since galaxy isophotes twist in the outer regions \citep[first observed by][]{Evans}. Since the mean tangential ellipticity (shown in Fig. \ref{fig:ep_wf}) depends on the ellipticity distribution, the observed alignment signal dependence could be caused by these differences. However, we have checked, using a different alignment estimator, the mean satellite alignment angle $\langle \phi_{\rm sat}\rangle$ (Fig. \ref{fig:angles}), that the behaviour is present. Therefore, the behaviour of Fig. \ref{fig:ep_wf} cannot be explained by the differences of ellipticity measurements between the different weight functions.

Recent measurements of alignments in groups and clusters have yielded conflicting results \citep{Schneider,Elisagroups,Sifon2015,Huang2}. Cluster studies have been performed using shape measurement methods that optimize the size of the weight function with SNR (such as the KSB or re-Gaussianization methods) and find no significant alignment \citep[e.g.][]{Elisagroups, Sifon2015}. We have tested this approach with DEIMOS and found that the highest SNR is obtained when the weight function is as small as possible, where the bulge dominates the light profile. Given that the satellite alignment signal drops for smaller weight functions (Figure \ref{fig:ep_wf}), studies using this approach are not expected to detect a strong alignment signal. In addition, the shape noise and statistics of these studies drive up the error bars (in comparison to our work) and make the detection even less likely. Model-fitting shape measurement methods have also been used in some of the studies above, but since the bulge of the galaxy carries most of the light profile it is also possible that the derived shape from such methods is greatly influenced by the inner galaxy regions, and the alignment signal is again hard to detect. In GAMA groups, \citet{Schneider} looked into satellite alignments with shapes from similar shape measurement methods applied to SDSS images and found no significant alignment, which can be accounted for in the same way as for the cluster studies.

In contrast to the studies described above, \citet{Huang2} looked at satellite alignments in clusters using three different shape estimators: moment-based, model-fitting and isophote fitting. They found a significant alignment that decreases for galaxies further away from the BGG, in agreement with out results. A difference in alignment was also observed for the three different estimators, which are sensitive to different galaxy scales. However, these estimators have different multiplicative biases, which were not corrected for, and understanding how much of the bias drives these differences is non-trivial.

\section{BGG shape - satellite position alignment}
\label{sec:BGGsat}

Intra-halo alignments can have multiple components, as one can correlate the shape or position of satellites with the shape or position of the central galaxy. In the previous section we have focused on the alignment of satellite shapes with the position of the central galaxy. Another interesting component, useful when constructing the halo model for intrinsic alignments, is the alignment of the central galaxy shape with the position of satellites. This is computed through Eq. \ref{eq:eplus} by substituting the ellipticity components with the ones of the BGG. 

\begin{figure}
	\resizebox{\hsize}{!}{\includegraphics{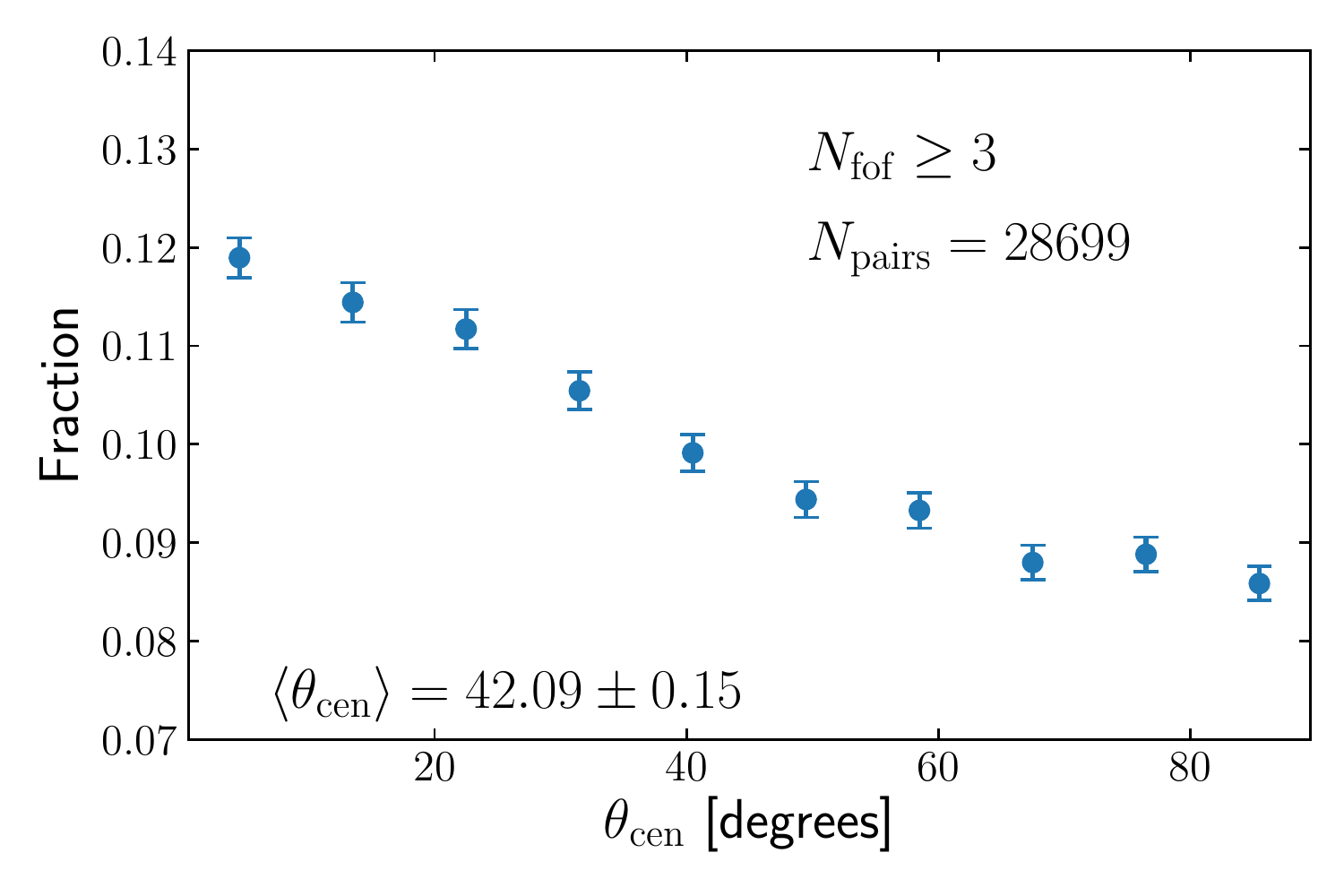}}
	\caption{Distribution of the angle $\theta_{\rm cen}$ between the BGG's semi-major axis and the line connecting the BGG-satellite pair. The mean value of the angle is shown in the bottom left of the plot.}
	\label{fig:thetacen}
\end{figure}

We compute the angle $\theta_{\rm cen}$ between the BGG's semi-major axis and the line connecting a pair of BGG-satellite galaxies. When this angle is 0$^\circ$ (or 90$^\circ$) the BGG is pointing radially (or tangentially) to the direction of the satellite. When $\langle \theta_{\rm cen}\rangle=45^\circ$ the satellites are randomly (circularly) positioned around the BGG but when this average angle is smaller than $45^\circ$, they are mostly distributed along the BGG's semi-major axis. In Figure \ref{fig:thetacen} we plot the fraction of BGG-satellite pairs against the value of $\theta_{\rm cen}$. We see that the distribution is not uniform, with more satellites positioned close to the BGG's semi-major axis rather than perpendicular to it. The average angle is computed to be $\langle\theta_{\rm cen}\rangle=42.09\pm0.15$, which is significantly lower from 45$^\circ$.

\begin{figure}
	\resizebox{\hsize}{!}{\includegraphics{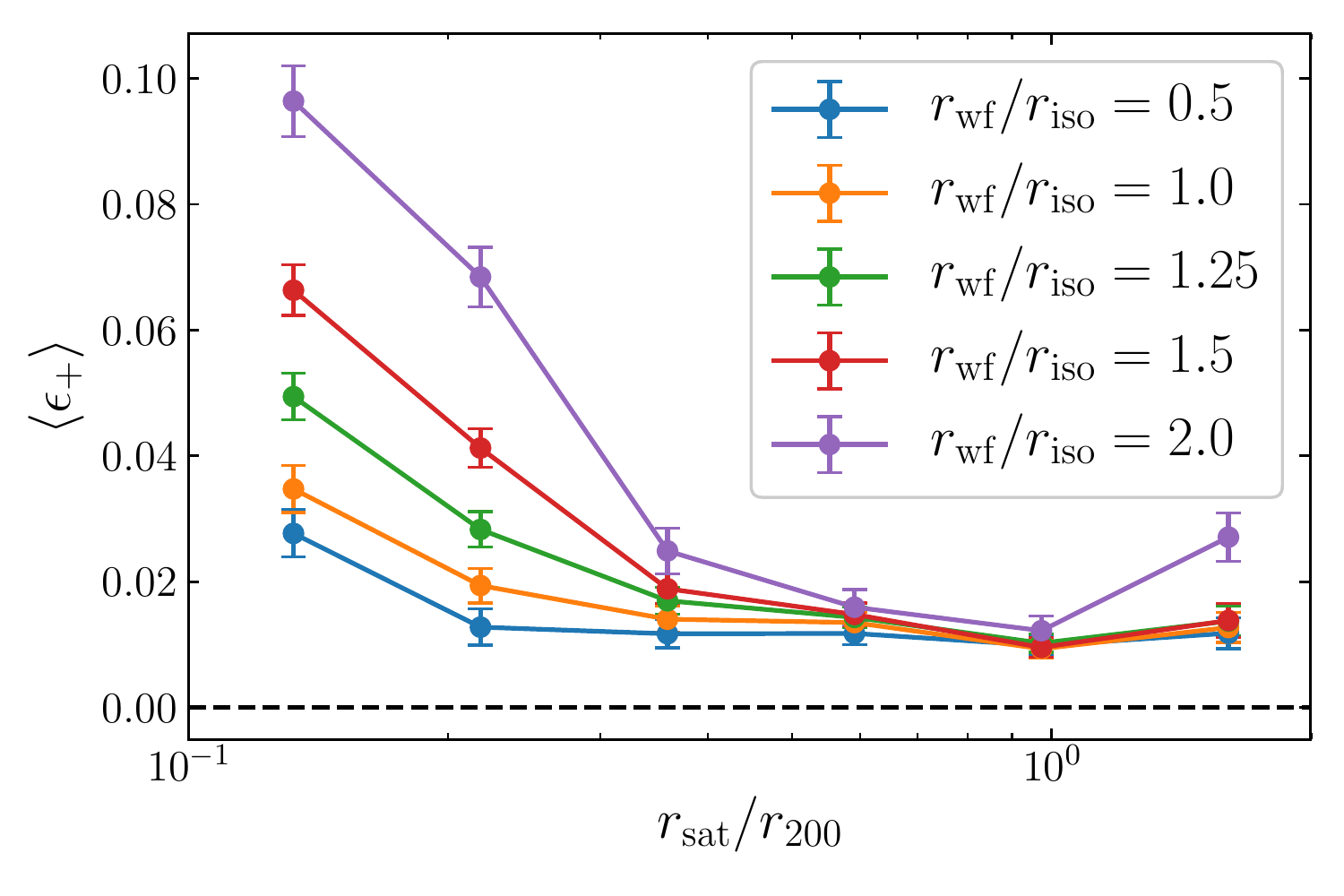}}
	\caption{Average tangential ellipticity of the BGG shape with respect to the satellites position, as a function of the projected distance of the satellite to the BGG. Results are obtained using 5 different weight function sizes, and are shown with a different colour.}
	\label{fig:BCGeplus}
\end{figure}

We also compute the average tangential ellipticity of the BGG's shape with respect to the satellite's position. This statistic is more easily related to the halo model formalism, and the mean $\epsilon_+$ is shown in Figure \ref{fig:BCGeplus} as a function of galaxy pair separation, for shapes obtained with the 5 different weight functions. The average tangential ellipticity is non-zero and higher for galaxy pairs closer together, dropping to a plateau at larger separations. In addition, the alignment is stronger for shapes obtained with a larger weight function. 

From this we conclude that satellites are preferentially distributed along the BGG's semi-major axis. Satellites closer to the BGG are more tightly aligned with this axis. Furthermore, the outer regions of the BGG are more aligned with the overall satellite distribution, as can be seen from the dependence of the alignment signal on the weight function size, particularly on smaller scales. Assuming that the satellite distribution traces the shape of the dark matter halo of the galaxy group, we conclude that the outer regions of the BGG are aligned more strongly with the group halo shape than the inner regions. This could be the result of the gravitational interaction of the BGG with the dark matter halo, or a reflection of the fact that streams of infalling matter in a galaxy group follow the ellipticity of the group's halo. 

\begin{figure}
	\resizebox{\hsize}{!}{\includegraphics{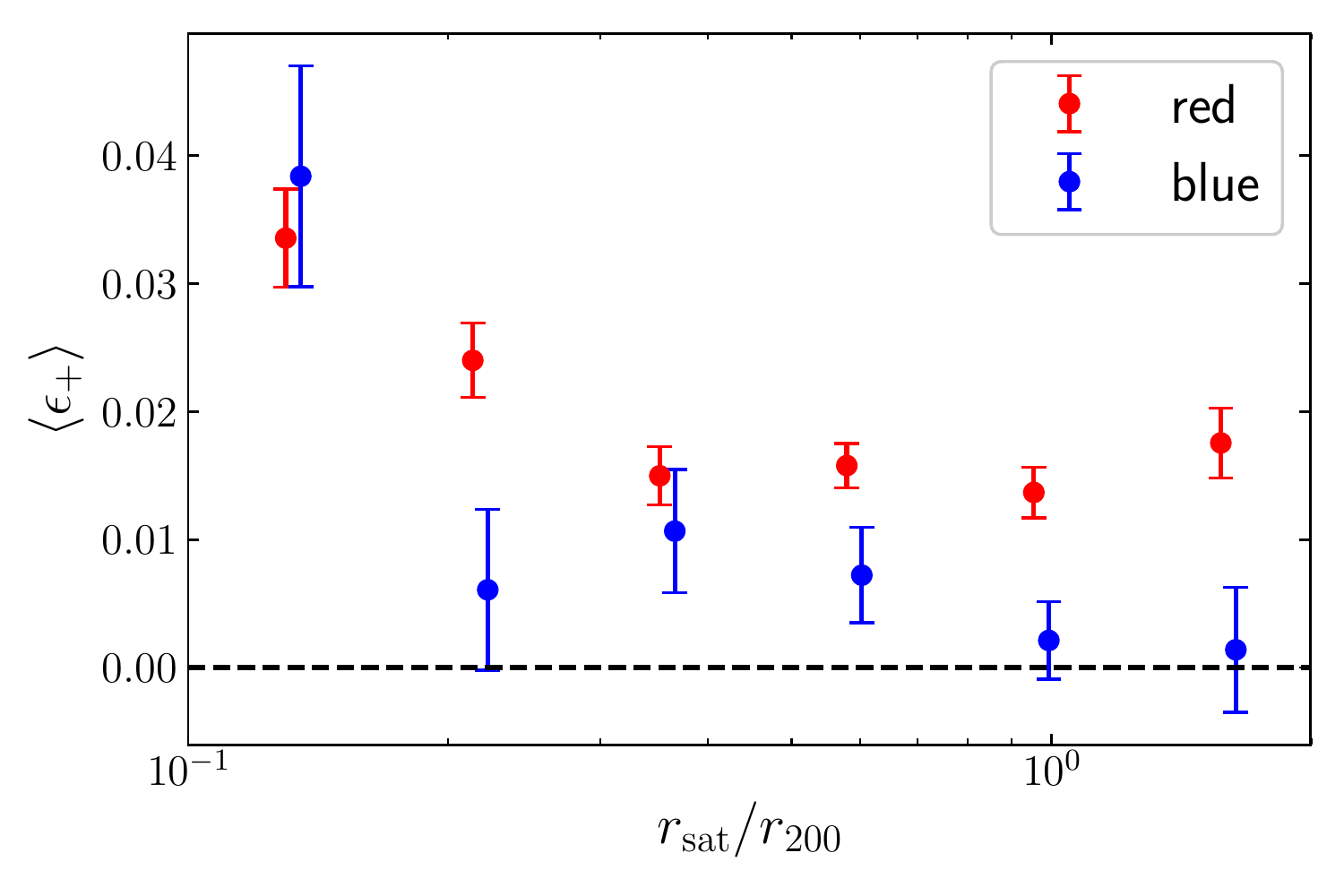}}
	\caption{Average tangential ellipticity of the BGG shape with respect to the satellites position, for red and blue BGGs. Results are obtained using $r_{\rm wf}=r_{\rm iso}$.}
	\label{fig:BCGeplus_redblue}
\end{figure}

We now split the BGGs into two populations: blue, high sSFR and red, low sSFR galaxies, using Eq. \eqref{eq:bluered}. This essentially splits the group samples into groups with a red or a blue central galaxy. We note here that 37\% of groups in our sample contain a blue BGG; these are likely groups that were formed recently, and are very different from galaxy clusters, where the brightest cluster galaxy will almost always be red. The alignment of the BGG shape with the satellite positions for these two sub-samples is shown in Figure \ref{fig:BCGeplus_redblue}. We see that the alignment is driven mostly by red BGGs, with blue BGGs being strongly aligned with their innermost satellites, weakly aligned in intermediate scales and unaligned with the positions of their outermost satellites. These results are in agreement with what has been found in recent literature \citep[e.g.][]{SDSSgroups, EdoHalo, IAenv}.

\section{Global intrinsic alignments}
\label{sec:centrals}

In the previous sections we saw that satellite-BGG and BGG-satellite alignments exhibit a galaxy scale dependence, with the outer galaxy regions aligned more strongly than the inner ones. This naturally leads to the question whether halo-halo alignments (among central galaxies) also exhibit this dependence. We therefore calculate the projected shape - position correlation function from a shape and density sample consisting of the non-satellite galaxies of the GAMA group catalogue, which we call central galaxies sample. The correlation function is computed from 
\begin{equation}
w_{g+}(\mathbf{r}_p)=\int_{-\Pi_{\max}}^{+\Pi_{\max}} \xi_{g+}(\mathbf{r}_p, \Pi)\mathrm{d}\Pi\,,
\label{eq:wg+theory}
\end{equation}
where $r_p$ and $\Pi$ is the transverse and line of sight separation between a galaxy pair, respectively, and $\xi_{g+}$ is the three dimensional position-shape correlation function, measured using the modified Landy-Szalay estimator \citep{LandySzalay, edo2017}
\begin{equation}
\xi_{g+}(\mathbf{r}_p, \Pi)=\frac{S_+D}{D_{\rm S}D}-\frac{S_+R}{D_{\rm S} R}\,.
\label{eq:xig+}
\end{equation}
In the above equation we define three galaxy samples, density ($D$), shape ($S$) and random points ($R$) with the density of the shape sample being $D_S$. The quantities $D_S D$ and $D_S R$ are normalized galaxy pairs \citep[see e.g.][for more details]{KirkReview} and 
\begin{equation}
S_+D = \sum_{i\neq j}\epsilon_+ (i|j)\,.
\label{eq:S+D}
\end{equation}
Galaxy $i$ is selected from the shape sample and $j$ from the density sample. Similarly we compute $S_+ R$, using the random catalogues specifically produced for GAMA galaxies \citep{Randoms}.

Following \citet{Georgiou}, we compute the difference in alignment 
\begin{equation}
\Delta w_{g+} = w_{g+}(r_{\rm wf}=2r_{\rm iso}) - w_{g+}(r_{\rm wf}=r_{\rm iso})\,,
\label{eq:dwgp}
\end{equation}
where the first term of the right hand side is computed with shapes obtained in the $r$-band, using a weight function $2r_{\rm iso}$ and the second term using $1r_{\rm iso}$, both from shape and density samples of GAMA central galaxies. If $\Delta w_{g+}$ is significantly positive then central galaxies also exhibit a similar scale dependence to the one seen from satellite galaxies in Section \ref{sec:scaledependence}.

To compute error bars for our measurements we run 100 realisations of the $\Delta w_{g+}$ measurement but adding a random position angle to galaxies each time. \citet{Georgiou} showed that error bars obtained this way agree with bootstrap estimated errors, while other techniques such as jackknife might not be optimal. We ensure that the same galaxies are used when computing the two $w_{g+}$ terms of \eqref{eq:dwgp} which ensures that the sample variance contribution to the error bar can be neglected, since we are subtracting signals affected by the same sample variance \citep[see][for a discussion]{Georgiou}.

\begin{figure}
	\resizebox{\hsize}{!}{\includegraphics{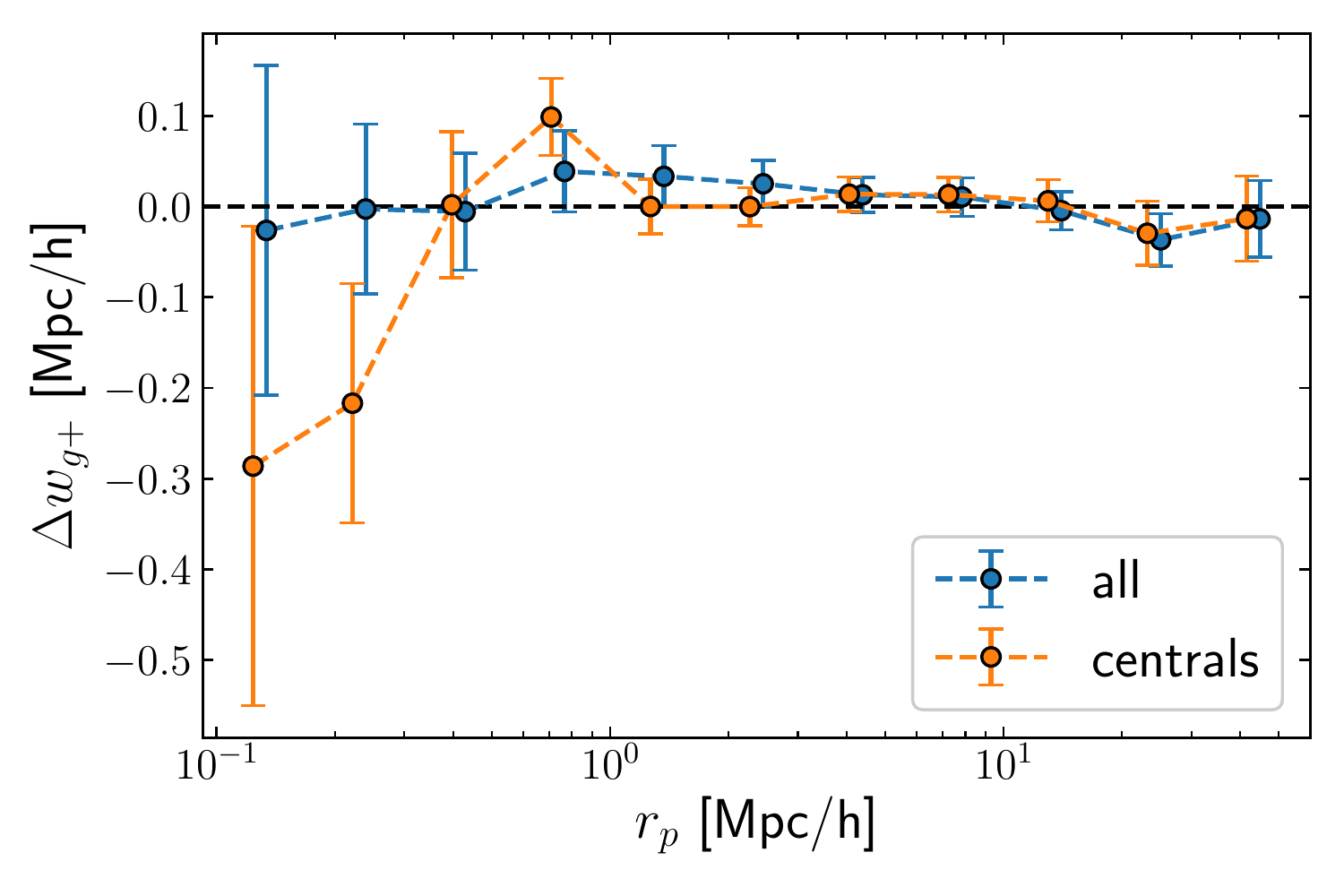}}
	\caption{Intrinsic alignment signal difference, as measured from shape measurements of two different weight functions, $r_{\rm wf}=\{r_{\rm iso}, 2r_{\rm iso}\}$, as a function of projected galaxy pair separation. Results obtained from shape and density samples consisting of central galaxies are shown with an orange dashed line, while the same results using all GAMA galaxies are shown with the blue solid line. Points are horizontally displaced for clarity.}
	\label{fig:wgp}
\end{figure}

Figure \ref{fig:wgp} shows the results for the alignment signal difference of central galaxies using two weight functions in the shape measurement. The alignment difference $\Delta w_{g+}$ is consistent with zero. It is clear that, for central galaxies, no galaxy scale dependence is evident, and the two alignment measurements agree well within the errorbars. In addition, we repeat the same exercise using the full GAMA galaxy sample and find differences consistent with zero, except for intermediate scales, $\sim1-2$ Mpc$/h$, where the alignment signal difference is consistently positive. However the error bars of our measurements do not allow for a robust detection. 

Studies of cosmological simulations have measured a dependence of the global intrinsic alignment signal on the way galaxy shapes were determined \citep[e.g.][]{Marco, Elisasim, Hilbert}. Shapes derived with a radial weighting scheme (therefore suppressing outer galaxy regions) revealed a lower amplitude of IA than shapes without a weight, which is not in agreement with our results. However, shapes that have been obtained with a weighting scheme must also suffer from shape bias, as well as bias due to the finite number of particles used to define galaxies. The effect of such bias is unclear. Furthermore, the intrinsic alignment amplitude measured in these simulations was found to be higher than the amplitude measured in GAMA \citep{Harry}, and the galaxy population in these simulations does not necessarily resemble the sample we used in this work, therefore a direct robust comparison is difficult. 

Comparing to other observations, \citet{Singh2} found, using low redshift luminous red galaxies, that alignment measurements with different shape estimators resulted in different inferred intrinsic alignment amplitudes. One of the possible reasons for this is that these shape estimators are sensitive to different galaxy scales. However, we do not find such a dependence in this work when we consider centrals or the full galaxy sample. This can be due to the fact that the sample of galaxies in the two studies are different, with the GAMA sample containing also fainter, blue galaxies with a wider redshift range. Indeed, the alignment signal of GAMA galaxies is measured to have a much smaller amplitude \citep{Harry} than the galaxies in \citet{Singh2}. Multiplicative bias can also cause such a difference, as the observed alignment signal scales linearly with bias \citep[Eq. 24][]{Singh2}, and the shape measurements in \citet{Singh2} were not corrected for it.

\section{Conclusion}
\label{sec:Conclusions}

We quantified the alignment of satellite galaxies with respect to the BGG in galaxy groups. Using galaxies from the GAMA survey and KiDS imaging data in $g,r$ and $i$-band filters we ensured a high fidelity galaxy group catalogue and accurate shape measurements from the DEIMOS shape measurement method. We controlled systematic effects due to the shape measurement process as well as contamination to the alignment signal from the extended light profile of the BGG with dedicated image simulations. 

The satellite alignment signal was obtained by calculating the average tangential ellipticity of satellite galaxies relative to the BGG position. Our measurements for shapes in the $r$-band revealed a significant radial satellite alignment signal that is stronger for satellites closer to the BGG and vanishes at large projected separations. Measurements in the $g$ and $i$-band images implied an apparent stronger alignment in these filters compared to $r$-band observations, but the data did not allow for a robust conclusion. We did not detect any significant trend of the amplitude of the satellite alignment signal as a function of satellite/BGG absolute magnitude or of group mass. 

We also split the satellite sample into red galaxies with low sSFR and blue galaxies with high sSFR. We found that red satellites have a higher alignment signal than blue satellites. This behaviour could arise from the fact that blue satellites have spent less time in the environment of the group, therefore having weaker alignment with it, as well as the difference in alignment mechanism between the two galaxy populations.

Furthermore we probed the galaxy scale dependence of satellite alignments. We varied the radial weight function employed during the shape measurement process, which results in shape measurements that reveal different parts of the galaxy. A small weight function gives more emphasis to the central region of a galaxy while shapes with a large weight function are also affected by the galaxy outer regions. Fitting a power-law function with a fixed slope of $-2$, we found a dependence of the satellite radial alignment on the weight function size for both red and blue satellite galaxies. The alignment signal is stronger for outer satellite galaxy regions. Notably, we detected a strong alignment signal for blue satellites around their BGG when using the two largest weight functions. These results are in line with the expectation that intrinsic alignments are mainly generated by tidal forces and outer, less bound, regions of a galaxy are more affected by these forces.

We found that satellites are preferentially distributed along the semi-major axis of the BGG, with the innermost satellites being more strongly aligned than the outermost. In addition, when the BGG's shape is determined with a larger weight funciton, the alignment at small scales is stronger, meaning that outer isophotes of the BGG are more strongly aligned with their satellites. Splitting the sample into groups with a red or a blue BGG, we found that red BGGs are more strongly aligned than blue ones, the latter being aligned mostly with their close satellites. It is not clear whether the BGG-satellite alignment scale dependence is caused by the tidal nature of the intrinsic alignments or from the direction of the in-falling matter into the galaxy group.

We also examined whether the global intrinsic alignment signal (correlating all galaxy shapes with all galaxy positions) depends on the galaxy scale. We computed the projected position-shape correlation function using shapes from two different weight functions and then subtracted the two alignment measurements. We found that the intrinsic alignment signal difference, when using only central galaxies, is consistent with zero; no galaxy scale dependence is detected for central galaxies. When we consider the whole GAMA galaxy sample there seems to be a positive difference of alignment signal in intermediate scales, which however is small compared to our error bars. From this result we conclude that the global intrinsic alignment signal does not depend strongly on the galaxy scale, within the errors of our measurements. 

The GAMA sample is flux limited and the galaxy population resembles samples commonly used in cosmic shear studies \citep[see e.g.][]{Harry}. Since the galaxy scale dependence on the alignment signal is measured to be consistent with zero, we conclude that weak gravitational lensing studies that use informative priors of intrinsic alignments, where the alignment signal is measured with a different shape estimator than the shear signal, will not be biased because of the difference in shape measurements.

\bibliographystyle{aa} 
\bibliography{references} 

\begin{acknowledgements}
	Based on data products from observations made with ESO Telescopes at the La Silla Paranal Observatory under programme IDs 177.A-3016, 177.A-3017 and 177.A-3018, and on data products produced by Target/OmegaCEN, INAF-OACN, INAF-OAPD and the KiDS production team, on behalf of the KiDS consortium. OmegaCEN and the KiDS production team acknowledge support by NOVA and NWO-M grants. Members of INAF-OAPD and INAF-OACN also acknowledge the support from the Department of Physics \& Astronomy of the University of Padova, and of the Department of Physics of Univ. Federico II (Naples). GAMA is a joint European-Australasian project based around a spectroscopic campaign using the Anglo-Australian Telescope. The GAMA input catalogue is based on data taken from the Sloan Digital Sky Survey and the UKIRT Infrared Deep Sky Survey. Complementary imaging of the GAMA regions is being obtained by a number of independent survey programmes including GALEX MIS, VST KiDS, VISTA VIKING, WISE, Herschel-ATLAS, GMRT and ASKAP providing UV to radio coverage. GAMA is funded by the STFC (UK), the ARC (Australia), the AAO, and the participating institutions. The GAMA website is http://www.gama-survey.org/.
	HH acknowledges support from Vici grant 639.043.512, financed by the Netherlands Organisation for Scientific Research (NWO). KK acknowledges support by the Alexander von Humboldt Foundation. HH acknowledges support from the European Research Council under FP7 grant number 279396 and the Netherlands Organisation for Scientific Research (NWO) through grants 614.001.103. NEC is supported by a Royal Astronomical Society research fellowship. MB is supported by the Polish Ministry of Science and Higher Education through grant DIR/WK/2018/12. CH and BG acknowledges support from the European Research Council under grant number 647112.
\end{acknowledgements}

\end{document}